\documentclass{aa}  

\usepackage{graphicx}
\usepackage{subfigure}
\usepackage{subfloat}
\usepackage{float}
\usepackage[version=3]{mhchem}
\usepackage{booktabs}
\usepackage{txfonts}
\begin{document} 

\title{Protoplanetary disk masses from CO isotopologues line emission}

%\subtitle{I. Overviewing the $\kappa$-mechanism}

\author{A. Miotello\inst{\ref{inst:leiden},\ref{inst:mpe}} \and S. Bruderer \inst{\ref{inst:mpe}}\and
        E. F. van Dishoeck\inst{\ref{inst:leiden},\ref{inst:mpe}}}
          
\institute{
Leiden Observatory, Leiden University, Niels Bohrweg 2, NL-2333 CA Leiden, The Netherlands\label{inst:leiden}
\and
Max-Planck-institute f{\"u}r extraterrestrische Physik, Giessenbachstra{\ss}e, D-85748 Garching, Germany\label{inst:mpe}
}
                       
%   \date{Received September 15, 1996; accepted March 16, 1997}

% \abstract{}{}{}{}{} 
% 5 {} token are mandatory
\abstract{One of the methods for deriving disk masses relies on direct observations of the gas, whose bulk mass is in the outer cold ($T\lesssim 30$K) regions. This zone can be well traced by rotational lines of less abundant CO isotopologues such as $^{13}$CO, C$^{18}$O and C$^{17}$O, that probe the gas down to the midplane. The total CO gas mass is then obtained with the isotopologue
ratios taken to be constant at the elemental isotope values found in the local ISM. This approach is however imprecise, because isotope selective processes are ignored.} % Context
{The aim of this work is an isotopologue selective treatment of CO isotopologues, in order to obtain a more accurate determination of disk masses.} % Aims
{The isotope-selective
photodissociation, the main process controlling the abundances of CO isotopologues in the CO-emissive layer,
is properly treated for the first time in a full disk model \citep[DALI,][]{Bruderer12,Bruderer13}. The chemistry, thermal balance, line and continuum radiative transfer are all
considered together with a chemical network that treats $^{13}$CO, C$^{18}$O, C$^{17}$O, isotopes of all included atoms, and molecules, as independent species. } % Methods
{Isotope selective processes lead to regions in the disk where the isotopologues abundance ratios e.g. of C$^{18}$O/$^{12}$CO are considerably different from the elemental $^{18}$O/$^{16}$O ratio. The results of this work show that considering CO isotopologue ratios as constants can lead to an underestimate of disk masses by up to an order of magnitude or more if grains have grown to larger sizes. This may explain observed discrepancies in mass determinations from different tracers. 
The dependence of the various isotopologues emission on stellar and disk parameters is investigated, to
set the framework for the analysis of ALMA data.} % Results
{Including CO isotope selective processes is crucial to determine the gas mass of the disk accurately (through ALMA observations) and thus to provide the amount of gas which may eventually form planets or change the dynamics of forming planetary systems.}% Conclusions

\keywords {}
%   {giant planet formation --
%                $\kappa$-mechanism --
%                stability of gas spheres
%               }

\maketitle
%
%________________________________________________________________

\section{Introduction}

Despite considerable progress in the field of planet formation in
recent years, many aspects are still far from understood
\citep[see][for review]{Armitage11}. What is clear is that the initial
conditions play an important role in the outcome of the planet
formation process. Circumstellar disks, consisting of dust and gas,
orbiting young stars are widely known to be the birth places of
planets. One of the key properties for understanding how disks evolve
to planetary systems is their overall mass, combined with their
surface density distribution.

So far, virtually all disk mass determinations are based on
observations of the millimeter (mm) continuum emission from dust
grains \citep[e.g.,][]{Beckwith90,Dutrey96,Mannings97,Andrews05}
\citep[see][for review]{Williams11}. To derive the total gas + dust
disk mass from these data involves several steps and
assumptions. First, a dust opacity value $\kappa_\nu$ at the observed
frequency $\nu$ together with a dust temperature needs to be chosen to
infer the total dust mass. The submillimeter dust opacity has been
calibrated for dense cores against infrared extinction maps
\citep{Shirley11} and found to be in good agreement with theoretical
opacities for coagulated grains with thin ice mantles
\citep{Ossenkopf94}, but the dust grains in protoplanetary disks have
likely grown to larger sizes with corresponding lower opacities at sub-mm wavelengths
\citep[e.g.,][]{Testi03,Rodmann06,Lommen09,Ricci10}. These opacities
may even vary with radial distance from the star, adding another
uncertainty \citep{Guilloteau11,Perez12,Birnstiel12} \citep[see][for
review]{Testi14}.  Second, a dust to gas mass ratio has to be assumed,
usually taken to be the same as the interstellar ratio of 100.  This
conversion implicitly assumes that the gas and mm-sized dust grains
have the same distribution. There is now growing observational
evidence that mm-sized dust and gas can have very different spatial
distributions in disks
\citep[e.g.,][]{Panic09,Andrews12,vanderMarel13,Bruderer14,Walsh14},
invalidating the use of mm continuum data to trace the gas.

The alternative method for deriving disk masses relies on observations
of the gas. The dominant constituent, H$_2$, is very difficult to
observe directly because of its intrinsically weak lines at near- and
mid-infrared wavelengths superposed on a strong continuum
\citep[e.g.,][]{Thi01,Pascucci06,Carmona08,Bitner08,Bary08}. Even if
detected, H$_2$ does not trace the bulk of the disk mass in most cases 
\citep[e.g.,][]{Pascucci13}. HD is a good alternative probe but its
far-infrared lines have so far been detected for only one disk
\citep{Bergin13} and there is no current facility sensitive enough for
deep searches in other disks after {\it Herschel}.

This leaves CO as the best, and probably only, alternative to
determine the gas content of disks.  In contrast with H$_2$ and HD,
its pure rotational transitions at millimeter wavelengths are readily
detected with high signal to noise in virtually all protoplanetary
disks
\citep[e.g.,][]{Dutrey96,Thi01,Dent05,Panic08,Williams14}. It is the second most
abundant molecule after H$_{2}$, with a chemistry that is in principle
well understood.  However,
$\rm ^{12}CO$ is not a good tracer of the
bulk of the gas mass because its lines become optically thick at
the disk surface. 
Less abundant CO isotopologues such as $^{13}$CO and C$^{18}$O have
more optically thin lines and as a consequence saturate deeper in the
disk, with C$^{18}$O probing down to the midplane
\citep{vanZadelhoff01,Dartois03}. Therefore, the combination of
several isotopologues can be used to investigate both the radial and
vertical gas structure of the disk.  With the advent of the Atacama
Large Millimeter/submillimeter Array
(ALMA\footnote{www.almaobservatory.org}), high angular resolution
observations of CO isotopologues in disks will become routine even for low
mass disks \citep[e.g.,][]{Kospal13}, allowing studies of the
distribution of the cold ($<$100 K) gas in disks in much more detail
than possible before. The ALMA data complement near-infrared
vibration-rotation lines of CO which probe mostly the warm gas
in the inner few AU of the disk
\citep[e.g.,][]{Najita03,Pontoppidan08,Brittain09,vanderPlas09,Brown13}.

The two main unknowns in the determination of the disk gas mass are the
CO-H$_{2}$ abundance ratio and the isotopologue ratios. 
In the simplest situation, the bulk of the volatile carbon (i.e., the
carbon that is not locked up refractory dust) is contained in
gas-phase CO leading to a CO/H$_2$ fractional abundance of $\sim
2\times 10^{-4}$, consistent with a direct observation of this
abundance in a warm dense cloud \citep{Lacy94}. The isotopologue
ratios are then usually taken to be constant at the $^{13}$C, $^{18}$O
and $^{17}$O isotope values found in the local interstellar medium
(ISM) \citep{WilsonRood94}.  In reality, two processes act to decrease
the CO abundance below its maximum value: photodissociation and
freeze-out. Photodissociation is effective in the surface layers of
the disk, whereas freeze-out occurs in the cold ($T_{\rm d}<$20 K)
outer parts of the disk at the midplane. Indeed, a combination of both
processes has been invoked to explain the low observed abundances of
CO in disks compared with H$_2$ masses derived from dust observations
\citep{Dutrey97,vanZadelhoff01,Andrews11}. An additional effect is
that the volatile carbon abundance and [C]/[O] ratio in the disk can
be different from that in warm clouds \citep{Oberg11,Bruderer12} and
affect the CO abundance. Indeed, a recent study by \citet{Favre13} of
the one disk, TW Hya, for which its mass has been determined
independently using HD far-infrared lines, finds a low C$^{18}$O
abundance and consequently a low overall carbon abundance, which they interpret as due to conversion of gas-phase CO to other hydrocarbons. These other carbon-bearing
species have a stronger binding energy to the grains than CO itself
and freeze-out rapidly preventing conversion back to CO \citep{Bergin14}.

Of the above processes, only photodissociation by ultraviolet (UV)
photons can significantly affect the abundance ratios of $^{12}$CO and
its isotopologues.  CO is one of only a few molecules whose
photodissociation is controlled by line processes, initiated by
discrete absorptions of photons into predissociative excited states,
and is thus subject to self-shielding
\citep{Bally82,VDB88,Viala88}. For a CO column density of about
$10^{15}$ cm$^{-2}$, the UV absorption lines saturate and the
photodissociation rate decreases sharply allowing the molecule to
survive in the interior of the disk \citep{Bruderer13}. Because the abundances of
isotopologues other than $^{12}$CO are lower, they are not
self-shielded until deeper into the disk. This makes photodissociation
an isotope-selective process, in particular for the rarer C$^{18}$O
and C$^{17}$O isotopologues. Thus, there should be regions in the disk
in which these two isotopologues are not yet shielded, but $^{12}$CO
and $^{13}$CO are, resulting in an overabundance of $^{12}$CO and
$^{13}$CO relative to C$^{18}$O and C$^{17}$O.  A detailed study of CO
isotope selective photodissociation incorporating the latest molecular
physics information has been carried out by \citet{Visser09} and
applied to the case of a circumstellar disk. A single vertical cut in
the disk was presented to illustrate the importance of isotope
selective photodissociation, especially when grains have grown to
larger sizes so that shielding by dust is diminished. If these effects
are maximal close to the CO freeze-out zone where most of the CO emission originates, the gas-phase emission
lines can be significantly affected. Other studies have considered
$^{13}$CO in disks but not the rarer isotopologues \citep{Willacy09}.

The aim of our work is to properly treat the isotope-selective
photodissociation in a full disk model, in which the chemistry,
gas thermal balance, and line and continuum radiative transfer are all
considered together. The focus is on the emission of the various
isotopologues and their dependence on stellar and disk parameters, to
set the framework for the analysis of ALMA data and retrieval of
surface density profiles and gas masses. In this first paper, we
present only a limited set of representative disk models to illustrate
the procedure and its uncertainties for the case of a disk around a T
Tauri and a Herbig Ae star. In \S 2, we present the model details,
especially the implementation of isotope selective processes. In \S 3,
the model results for our small grid are presented and the main
effects of varying parameters identified. Finally, in \S 4, the model
results and their implications for analyzing observations are
discussed. In particular, the case of TW Hya is briefly discussed.

%________________________________________________________________
\section{Model}
\label{model}

For our modeling, we use the  DALI (Dust And LInes) code \citep[][]{Bruderer12,Bruderer13}, 
a radiative transfer, chemistry and thermal-balance model. Given a density 
structure as input, the code solves the continuum radiative transfer using a 3D Monte
Carlo method to calculate the dust temperature $T_{\rm dust}$ and 
local continuum radiation field from UV to mm wavelengths.  A chemical network 
simulation then yields the chemical composition of the gas.  
The chemical abundances enter a non-LTE excitation calculation of the main 
atoms and molecules. The gas temperature $T_{\rm gas}$ is then obtained from the 
balance between heating and cooling processes. Since both the chemistry 
and the molecular excitation depend on $T_{\rm gas}$, the problem is 
solved iteratively. Once a self-consistent solution is found, spectral 
image cubes are created with a raytracer. The DALI code has been tested 
with benchmark test problems \citep[][]{Bruderer12,Bruderer13} and against 
observations \citep[][]{Bruderer12,Fedele13,Bruderer14}.

In this work, we have extended DALI with a complete treatment of isotope-selective 
processes. This includes a chemical network with different isotopologues taken as 
independent species (e.g. $^{12}$CO, $^{13}$CO, C$^{18}$O, and C$^{17}$O) and reactions which enhance or decrease the abundance of one isotopolog over the other.

\subsection{Isotope-selective processes}

The isotope selective processes included in the model are CO photodissociation and 
gas-phase reactions exchanging isotopes between species (fractionation reactions). 

\subsubsection*{Isotope-selective photodissociation}
\label{iso_sel_pd}
The main isotope-selective process in the gas phase is CO photodissociation 
\citep[][and references therein]{Visser09}. CO is photodissociated through discrete 
(line-) absorption of UV photons into predissociative bound states. Absorption of 
continuum photons is negligible. Since the dissociation energy of CO is 11.09 eV, 
CO photodissociation can only occur at wavelengths between 911.75 \AA~ and 1117.8 \AA. 
The UV absorption lines are electronic transitions in vibrational levels of excited states and can become 
optically thick. Thus, CO can shield itself from photodissociating photons. In particular 
the UV absorption lines of the main isotopologue $^{12}$CO become optically thick at a 
CO column density $\sim$$10^{15} \rm cm^{-2}$ \citep{VDB88}. In disks, this column density 
corresponds to the surface of the warm molecular layer. At a certain height in the disk, 
the photodissociation rate has dropped sufficiently for CO to survive, both due to self-
shielding and absorption of FUV continuum attenuation by small dust grains or PAHs.

The rarer isotopologues (e.g. $^{13}$CO, C$^{18}$O, and C$^{17}$O) can also self-shield 
from the dissociating photons, but at higher column densities and accordingly closer to 
the mid-plane. This results in regions where $^{12}$CO is already self-shielded and thus 
at high abundance, but the rare isotopologues are still photodissociated due to their 
less efficient self-shielding. In those regions the isotopologue ratio (e.g. C$^{18}$O/
$^{12}$CO) can be much lower than the corresponding elemental isotope ratio [$^{18}$O]/[$^{16}$O]. In 
chemical models of disks, the abundance of the rare isotopologues is usually obtained by 
simply scaling the $^{12}$CO abundance with the local ISM elemental isotope ratio \citep{WilsonRood94}. However, to correctly deduce the total gas disk mass from rare 
CO isotopologues observations, isotope-selective photodissociation needs to be taken into account.

The depth-dependence of the photodissociation rates is affected by different effects. 
Besides self-shielding, blending of UV absorption lines with other species can lead to 
mutual shielding. For example rare CO isotopologues can be mutually shielded by 
$^{12}$CO, if their UV absorption lines overlap. Also mutual shielding by H and H$_2$ is 
important. At larger depths, the UV continuum radiation is attenuated by small dust 
grains and PAHs. The photodissociation rate for a particular isotopologue $^x$C$^y$O can 
be written as
\begin{equation}
k_{\rm PD} = \Theta\left[ N({\rm H}), N({\rm H}_2), N({}^{12}{\rm CO}), N({}^x{\rm C}{}^y{\rm O})\right]\, k^0_{\rm PD} \ ,
\label{kappa_pd}
\end{equation}
where $\Theta$ is a shielding function depending on the H, H$_2$, $^{12}$CO and 
$^x$C$^y$O column densities and $k^0_{\rm PD} $ is the unshielded photodissociation 
rate, calculated using the local continuum radiation field. 

In this work, the mutual and self-shielding factors $\Theta$ are interpolated from the 
values given by \cite{Visser09}. We use their tables with intrinsic line widths 
$b_{\rm CO}=0.3$ km s$^{-1}$, $b_{{\rm H}_2}=3$ km s$^{-1}$, and $b_{\rm H}=5$ km s
$^{-1}$, and excitation temperatures $T_{\rm ex, CO}= 20$ K and 
$T_{\rm ex, {\rm H}_2}= 89.4$ K, values appropriate for the lower-$J$ lines. The column densities of H, H$_2$, $^{12}$CO and 
$^x$C$^y$O are calculated as the minimum of the inward/upward column density from the 
local position. This approach has been verified in \cite{Bruderer13} against the method 
used in \cite{Bruderer12}, where the column densities are calculated together with the 
FUV radiation field by the Monte Carlo dust radiative transfer calculation 
\citep[see Appendix A in][]{Bruderer13}. This method is computationally far less demanding, as it does not require a global iteration of the model.

\subsubsection*{Fractionation reactions} \label{sec:fractionation}

Besides isotope selective photodissociation, also gas-phase reactions can change the 
relative abundance of isotopologues. The most important reaction of this type is the 
ion-molecule reaction
\begin{equation}
\rm ^{13}C^+ +\ce{^{12}_{}C}O \, \leftrightarrows \rm \, \ce{^{13}_{}C}O + \ce{^{12}_{}C^+} + 35 \,\rm K
\label{frac_reac}
\end{equation}
\citep[see][and references therein]{Watson76,WoodsWillacy09}. 
The vibrational ground state energy difference of $^{12}$CO and $^{13}$CO corresponds to 
a temperature of 35 K. Thus, at low temperature, the forward direction is preferred 
leading to an increased abundance of $^{13}$CO relative to $^{12}$CO \citep{Langer84}. At 
high temperature, forward and backward reactions are balanced and the $^{12}$CO/$^{13}$CO 
ratio is not altered by the reaction. Following \cite{Langer84}, the 
reaction rate coefficient for the backward reaction is 
$k = \alpha (T/300 \rm \, K)^{\beta} \exp(-\gamma/T)$, where $\alpha=4.42\times10^{-10}$ 
cm$^3$ s$^{-1}$, $\beta= -0.29$ and $\gamma = 35$ K. For the forward reaction, the 
exponential factor is dropped ($\gamma = 0$).

Another isotope-exchange reaction considered in our work is 
\begin{equation}
\rm H^{12}CO^+ + \ce{^{13}_{}C}O \, \leftrightarrows \rm \, H^{13}CO^+ + \ce{^{12}_{}C}O + 9 \,\rm K \ .
\label{frac_reac2}
\end{equation}
Rate coefficients for this reaction have been measured by \cite{SmithAdams80}. Due to 
the small vibrational ground state energy difference, corresponding to temperatures lower 
than the CO freeze-out temperature, it has however only a minor impact in disks.

\subsection{Chemical Network}
\label{chem_net}

For our models, the list of chemical species by \cite{Bruderer12} is extended to include 
different isotopologues as independent species. By adding the elements, $^{13}$C, 
$^{17}$O and $^{18}$O, in addition to H, He, $^{12}$C, N, $^{16}$O, Mg, Si, S and Fe, the 
number of species is increased from 109 to 276 (see Table \ref{tab:chemspec}). We account 
for all possible permutations. For example CO$_2$ is expanded into 12 independent species 
($^{12}$C$^{16}$O$_2$, $^{12}$C$^{16}$O$^{17}$O, $^{12}$C$^{17}$O$_2$, \ldots). 
 
Our chemical reaction network is based on the UMIST 06 network \citep{Woodall07}. It is 
an expansion of that used by \cite{Bruderer12} and \cite{Bruderer13} to include reactions 
between isotopologues. The reaction types included are: H$_2$ formation on dust, freeze-
out, thermal and non-thermal desorption, hydrogenation of simple species on ices, gas-
phase reactions, photodissociation, X-ray and cosmic-ray induced reactions, PAH/small 
grain charge exchange/hydrogenation and reactions with H$^*_2$ (vibrationally excited H$_2$). The details of the implementation of these reactions are described in the Appendix A.3.1 of \cite{Bruderer12}. Specifically, a binding energy of 855 K was used for pure CO ice \citep{Bisschop06}. The elemental abundances are also the same as in \cite{Bruderer12}. The isotope ratios are taken to be  [$^{12}$C]/[$^{13}$C]=77, [$^{16}$O]/[$^{18}$O]=560 and [$^{16}$O]/[$^{17}$O]=1792 \citep{WilsonRood94}.

Reactions involving the rarer isotopologues are duplicates of those involving the dominant isotopologue. For example the reaction H + O$_2 \, \rightarrow$ OH + O is expanded to the 
reactions
\begin{subequations}
\begin{align*}
        \rm H + \ce{^{16}_{}O}\ce{^{18}_{}O}&\, \rightarrow \, \rm ^{16}OH + \ce{^{18}_{}O}\\
        \rm H + \ce{^{16}_{}O}\ce{^{18}_{}O}&\, \rightarrow \, \rm ^{18}OH + \ce{^{16}_{}O}\\
        \rm H + \ce{^{18}_{}O2}&\, \rightarrow \, \rm ^{18}OH + \ce{^{18}_{}O}\\
        \rm H + \ce{^{16}_{}O}\ce{^{17}_{}O}&\, \rightarrow \, \rm ^{16}OH + \ce{^{17}_{}O}\\
        \rm H + \ce{^{16}_{}O}\ce{^{17}_{}O}&\, \rightarrow \, \rm ^{17}OH + \ce{^{16}_{}O}\\
        \rm H + \ce{^{17}_{}O2}&\, \rightarrow \, \rm ^{17}OH + \ce{^{17}_{}O}\\
        \rm H + \ce{^{18}_{}O}\ce{^{17}_{}O}&\, \rightarrow \, \rm ^{18}OH + \ce{^{17}_{}O}\\    
	\rm H + \ce{^{18}_{}O}\ce{^{17}_{}O}&\, \rightarrow \, \rm ^{17}OH + \ce{^{18}_{}O}
\end{align*}
\end{subequations}
This procedure increases the number of reactions included in the network from 1463 to 9755. 
Since most of the rate coefficients for such reactions involving isotopologues are unknown, we 
follow the procedure by \cite{Roellig13} and assume that the isotopologue reactions with the 
same reactants, but different products, have the original reaction rate divided by the number 
of out-going channels. This implies assigning an equal probability to all branches. 

The fractionation reactions (Eq. \ref{frac_reac} and \ref{frac_reac2}) have also been added to 
the network. In addition to $^{13}$C we also add the analogous reactions for $^{18}$O and 
$^{17}$O. When available, the reaction rate coefficients are taken from \cite{Roellig13}, 
which are based on \cite{Langer84}. Coefficients for reactions with $^{17}$O are not provided 
and extrapolations are made. For reaction (\ref{frac_reac}), we use the 
same $\alpha$ and $\beta$ for C$^{17}$O as for $^{13}$CO and C$^{18}$O (Sect. 
\ref{sec:fractionation}). The value of $\gamma$ is obtained from scaling with the reduced mass 
of C$^{17}$O following the procedure described in \cite{Langer84}. For reaction
(\ref{frac_reac2}), the $\alpha$ and $\beta$ coefficients are taken to be the same as that with $^{18}$O. For $\gamma$, we assume the value to be the mean of the values 
for $^{16}$O and $^{18}$O, since the mass of $^{17}$O is the median of the $^{16}$O and 
$^{18}$O masses. For the conditions in protoplanetary disks, reaction (\ref{frac_reac2}) is less important for chemical fractionation than reaction (\ref{frac_reac}); hence, this simple assumption is appropriate.

\subsection{Parameters of the disk model}

For our modeling, we assume a simple parameterized density structure, similar to 
\cite{Andrews11}. Assuming a viscously evolving disk, where the viscosity $\nu = R^{\gamma}$ is a power-law of 
the radius \citep{Lynden-BellPringle74,Hartmann98}, the surface density is given by
\begin{equation}
\Sigma_{\rm gas}=\Sigma_c \,\left( \frac{R}{R_c} \right) ^{-\gamma} \exp \, \left[ - \,\left( \frac{R}{R_c} \right) ^{2-\gamma} \right] \ ,
\end{equation}
with a characteristic radius $R_c$ and a characteristic surface density $\Sigma_c$. The characteristic radius is 
fixed to $R_c=200$ AU and the characteristic surface density $\Sigma_c$ adjusted to yield a total disk 
mass $M_{\rm gas}$ with the outer radius fixed to $R_{\rm out}=400$ AU . The gas-to-dust ratio is assumed to be 100.

The vertical density structure follows a Gaussian with scale height angle 
$h=h_{\rm c} (R/R_{\rm c})^{\psi}$. For the dust settling, two populations of grains are 
considered, small (0.005 - 1 $\mu$m) and large (1 - 1000 $\mu$m) \citep{Dalessio06}. The 
scale height is $h$ for the the small grains and $\chi h$ for the large ones, where 
$\chi < 1$. The distribution of the surface density of the two species is given by the factor 
$f_{\rm large}$ as  $\Sigma_{\rm dust}=f_{\rm large}\, \Sigma_{\rm large} + (1-f_{\rm large})\Sigma_{\rm small}$. 

\begin{figure*}
\centering
             {\includegraphics[width=0.8\textwidth]{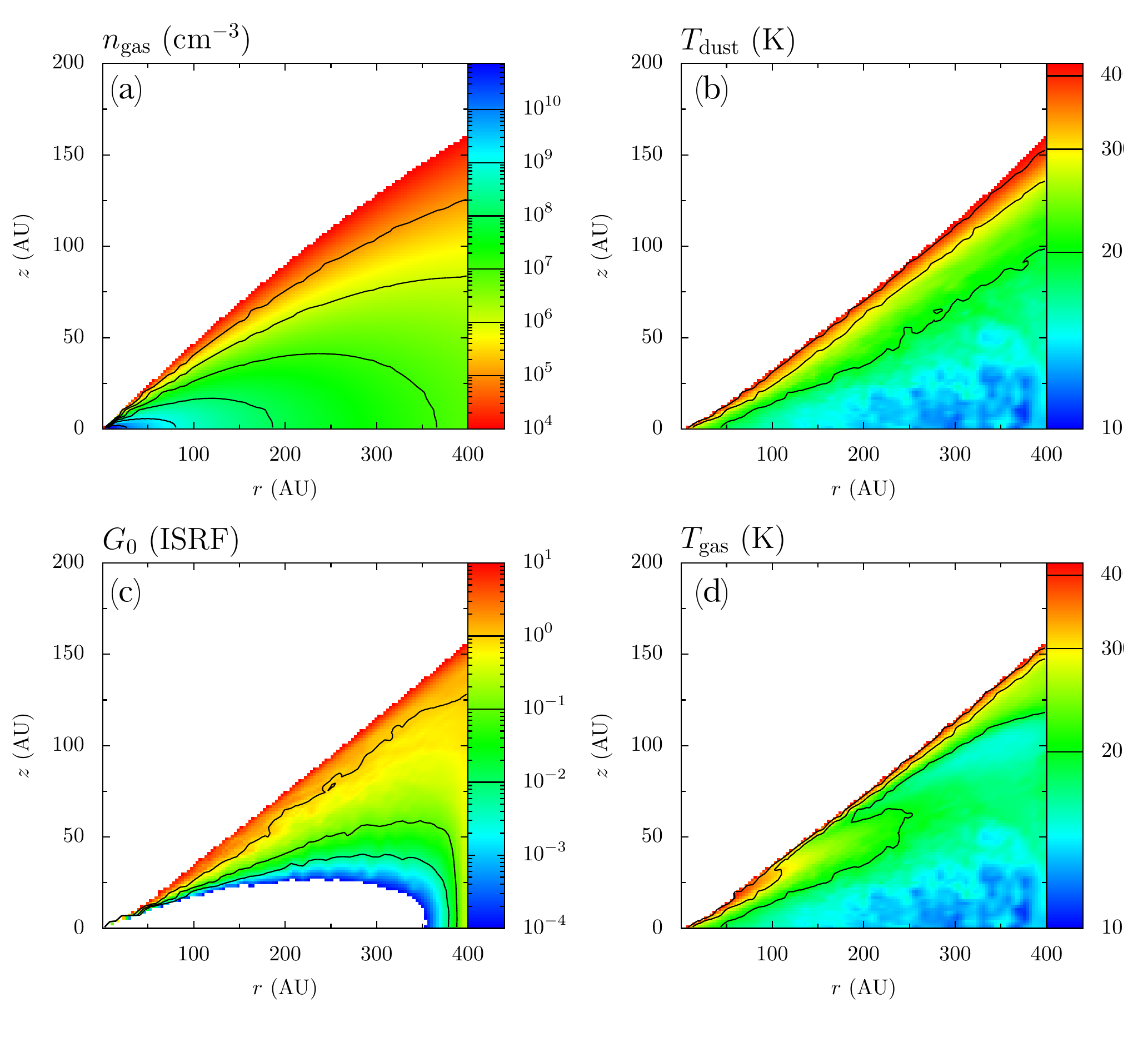}}
      \caption{2D representations of the results obtained including isotope-selective effects, for a representative model ($M_{\rm d}=10^{-2} M_{\odot}$, T Tauri star, $f_{\rm large}=10^{-2}$). The gas number density (panel a), the dust temperature (panel b), the FUV flux in units of $G_0$ (panel c) and the gas temperature (panel d) are shown. }
       \label{struc}
\end{figure*}

Other parameters of the model are the stellar FUV (6 -13.6 eV) and X-ray spectrum and the 
cosmic-ray ionization rate. In order to study the effects of different amounts of FUV photons 
compared with the bolometric luminosity, the stellar spectrum is assumed to be a black-body at 
a given temperature $T_{\rm eff}$. 
The strength of the FUV field in units of the interstellar
radiation field G$_0$ is given in Fig. \ref{struc} for a representative model. Here, $G_0$ = 1 refers
to the interstellar radiation field defined as in \cite{Draine78}
$\sim 2.7 \cdot 10^{-3} \rm erg \, s^{-1} \, cm^{-2}$ with photon-energies E$_{\gamma}$ between 6 eV
and 13.6 eV.
The X-ray spectrum is taken to be a thermal spectrum of $7 
\times 10^7$ K within 1 - 100 keV and the X-ray luminosity in this band $L_{\rm X} = 10^{30}$ 
erg s$^{-1}$. As discussed in \cite{Bruderer13}, the X-ray luminosity is of minor importance 
for the intensity of CO pure rotational lines which are the focus in this work. The cosmic-ray 
ionization rate is set to $5 \times 10^{-17}$ s$^{-1}$. We account for the interstellar UV 
radiation field and the cosmic microwave background as external sources of radiation.

The calculation is carried out on 75 cells in the radial direction and 60 in the vertical directions. In the radial direction the spatial grid is on a logarithmic scale up to 30 AU (35 cells) and on a linear scale from 30 AU to 400 AU (40 cells). The 
spectral grid of the continuum radiative transfer extends from 912 \AA $\,$ to 3 mm in 58 wavelength-bins. The wavelength dependence of the cross section is taken into account using data summarized in \cite{VD06}.

The chemistry is solved in a time dependent manner, up to a chemical age of 1 Myr. The 
main difference to the steady-state solution is that carbon is not converted into methane 
(CH$_4$) close to the mid-plane, since the time-scale of these reactions is longer than $>10$ 
Myr and thus unlikely to proceed in disks. We run models both with isotopologues and 
isotope-selective processes switched on (network \emph{ISO}) or off (network \emph{NOISO}).

\subsection{Grid of models}
\label{grid}

The goal of our work is to understand the effect of isotope-selective processes in disks and to 
quantify their importance when rare isotopologue observations are used to measure the total 
gas mass. We run a small grid of models to explore some of the key parameters:
\begin{enumerate}
\item \emph{Stellar spectrum}. The first parameter to explore is the fraction of FUV photons (912-2067 \AA)
over the entire stellar spectrum. A spectrum with $T_{\rm eff}=4000$ K and $T_{\rm eff}=10000$ K 
is considered, in order to simulate the case of a T Tauri or Herbig Ae star. Accordingly, we 
run models with a bolometric luminosity of $L_{\rm bol}=1$ or 10 $L_\odot$. Excess UV radiation due to accretion is taken into account in the case of T Tauri stars. 
It is assumed that the gravitational potential energy of accreted mass is released with 100$\%$ efficiency as blackbody emission with T=10000 K. The mass accretion rate is taken to be 10$^{-8} M_{\odot}$ yr$^{-1}$ and the luminosity is assumed to be emitted uniformly over the stellar surface. These assumptions result in $L_{\rm FUV}/L_{\rm bol}$= 1.5$\cdot10^{-2}$ for the T Tauri case versus $L_{\rm FUV}/L_{\rm bol}$= 7.8$\cdot10^{-2}$ for the Herbig case. The ratio of CO photodissociating photons (912-1100 \AA) between the T Tauri and the Herbig cases is $F_{\rm CO, pd}$(T Tau)/$F_{\rm CO, pd}$(Her) = 2.5$\cdot 10^{-2}$. In all models, the
interstellar UV field is included as well.

\item \emph{Dust properties}. Since small grains are much more efficient in absorbing UV 
radiation, the ratio of large grains to small grains is varied. We consider $f_{\rm large}=0.99$ in 
order to simulate a mixture of the two populations and $f_{\rm large} = 10^{-2}$ for the situation where 
only small grains are present in the disk (no or little grain growth).
\item \emph{Disk mass}. Three different values for the total disk mass $M_{\rm d}$ are 
considered in order to cover a realistic range of disks ($M_{\rm gas} = 10^{-2},  10^{-3},  10^{-4} M_{\odot}$).
\end{enumerate}

\begin{table}[tbh]
\caption{Parameters of the disk models.}
\label{tab:modelpar}
\centering
\begin{tabular}{ll}
\hline\hline
Parameter	 & Range \\
\hline
\emph{Chemistry} \\
Chemical network & ISO / NOISO \\
Chemical age & 1 Myr \\
\emph{Physical structure}& \\
$\gamma$ & 1\\
$\psi$ & 0.1 \\
$h_{\rm c}$ & 0.1 rad \\
$R_{\rm c}$ & 200 AU \\
$R_{\rm out}$ & 400 AU \\
$M_{\rm gas}$ & $10^{-4}$,\,$10^{-3}$,\,$10^{-2}$ \, M$_\odot$ \\
Gas-to-dust ratio & 100\\
$f_{\rm large}$ & $10^{-2}$, 0.99\\
$\chi$ & 1 \\
\emph{Stellar spectrum}&\\
T Tauri$^*$&$T_{\rm eff}$=4000 K, $L_{\rm bol}$=1 $L_{\odot}$\\
Herbig&$T_{\rm eff}$=10000 K, $L_{\rm bol}$=10 $L_{\odot}$\\
$L_{\rm X}$ & $\rm 10^{30}\, erg\,s^{-1}$\\
\emph{Dust properties}&\\
Dust & 0.005-1 $\mu$m (small)\\
& 1-1000 $\mu$m (large)\\
\hline
\end{tabular}
\newline
$^*$ \small{FUV excess added, see text.}
\end{table}

\begin{figure*}
\centering
             {\includegraphics[width=0.8\textwidth]{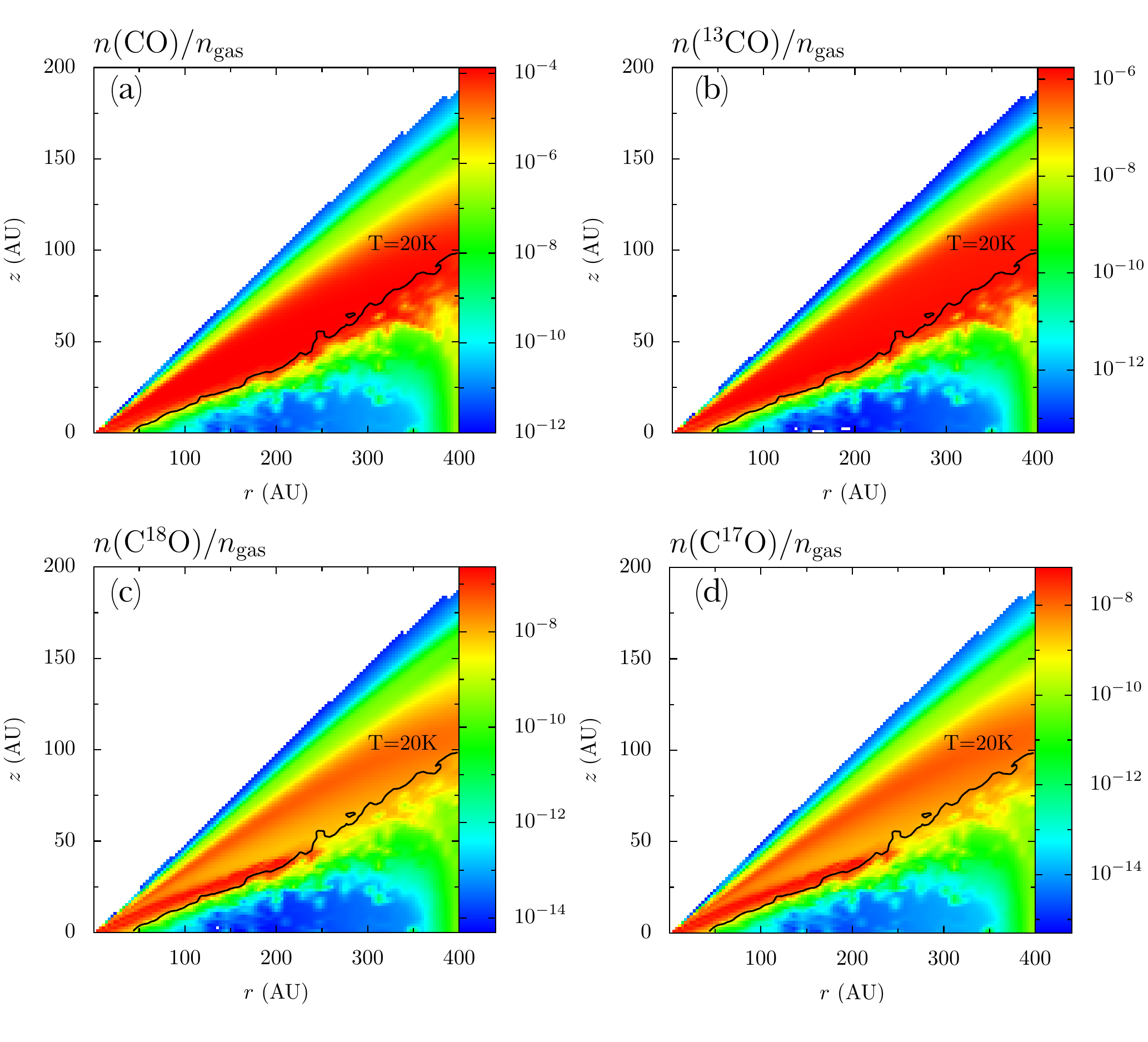}}
      \caption{2D representations of the results obtained including isotope-selective effects, for a representative model ($M_{\rm d}=10^{-2} M_{\odot}$, T Tauri star, $f_{\rm large}=10^{-2}$). The CO isotopologues abundances normalized to the total gas density are presented. The black solid line indicates the layer where the dust temperature is equal to 20 K. For lower T$_{\rm dust}$ values, CO freeze-out may become important.}
       \label{abund}
\end{figure*}

\begin{figure*}
   \resizebox{\hsize}{!}
             {\includegraphics[width=1.\textwidth]{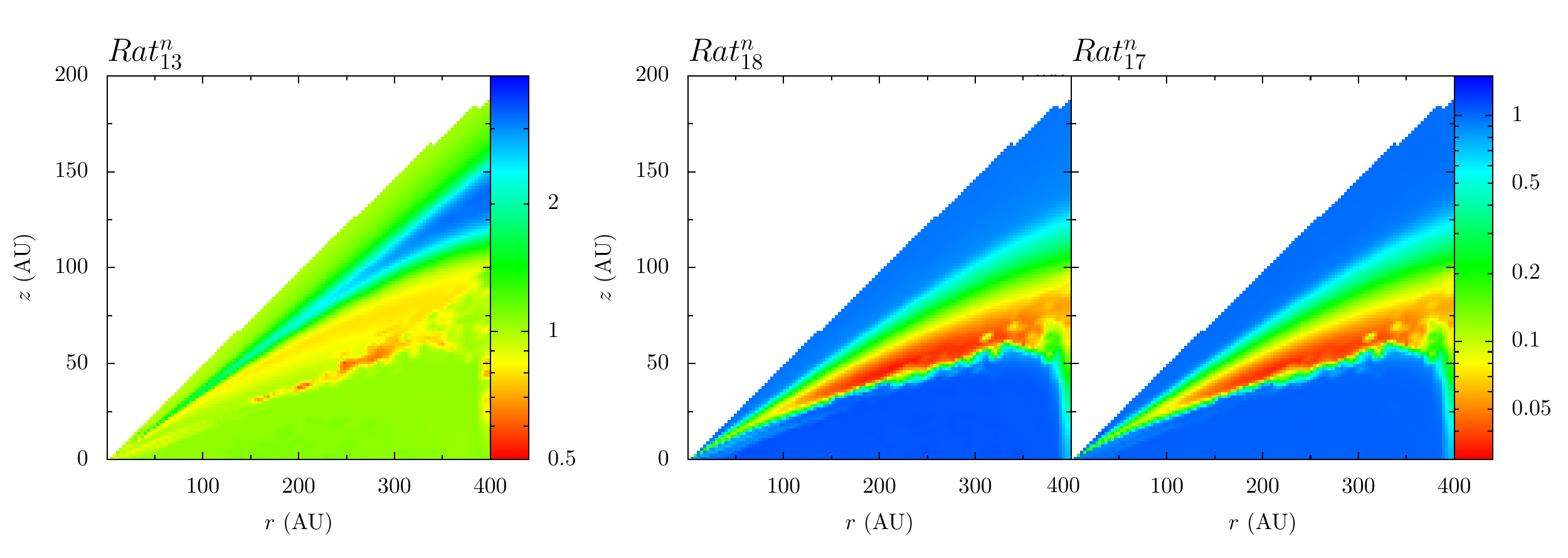}}
      \caption{CO isotopologues abundance ratios obtained using the NOISO and the ISO chemical networks. The ratios are defined as follows: Rat$^n_{13}= 77 \cdot \frac{n(\ce{^{13}_{}CO})_{\rm ISO}}{n(\rm CO)_{\rm NOISO}}$, Rat$^n_{18}= 560 \cdot \frac{n(\ce{C^{18}_{}O})_{\rm ISO}}{n(\rm CO)_{\rm NOISO}}$ and Rat$^n_{17}= 1792 \cdot \frac{n(\ce{C^{17}_{}O})_{\rm ISO}}{n(\rm CO)_{\rm NOISO}}$.
 The plots show the results for the case of a $10^{-2} M_{\odot}$ disk around a T Tauri star, with $f_{\rm large}=10^{-2}$. }
       \label{abu_ratio}
\end{figure*}

%____________________________________________________________________
\section{Results}
\label{results}

\subsection{Abundances}
In Figure \ref{abund} we present the abundances of CO and its isotopologues computed using the ISO network and implementing the isotopologue shielding factors in a representative model (T Tauri star, $M_{\rm disk}=10^{-2} M_{\odot}$, $f_{\rm large}=10^{-2}$.). 
In panel (a) a 2D distribution of $\ce{^{12}_{}CO}$ in a disk quadrant is shown; the central star is located at the origin of the axis. As the disk surface is directly illuminated by the FUV stellar and interstellar radiation, CO is easily photodissociated and its abundance drops, such that $n(\text{CO})/n_{\rm gas}$ reaches values around $10^{-8}$. This so-called photodissociation layer results in a low-abundance region shown in green in panel (a) of Figure \ref{abund}. The depth of this layer is not radially constant because it depends on the disk surface density structure and on the FUV flux. At large radii both $\rm H_2$ and CO column densities are lower and the FUV stellar flux is less strong: CO is not shielded until lower heights, compared to regions at smaller radii. As soon as the UV absorption lines saturate due to (self-) shielding, the photodissociation rate decreases steeply, allowing CO to survive in the interior of the disk. This warm, high-abundance layer of survived CO is shown in red in panel (a) of Figure \ref{abund}. It extends from the surface up to the midplane at very small radii, $R<40$ AU, as the column densities there are high enough that CO  can survive the strong stellar photodissociating flux, and the dust grains warm enough to prevent freeze-out. At radii larger than 40 AU, the abundance of CO falls again at decreasing height toward the midplane. There the dust temperature is very low ($T_{\rm dust} \lesssim 20 \,\rm K$) because the UV stellar flux is well shielded by the above regions. At such temperatures CO molecules freeze out onto dust grains, therefore the abundance of gaseous CO decreases. The precise dust temperature at which most of the CO is frozen out depends on the density and can be as low as 18 K for the adopted CO binding energy.

The shape of the distribution of $\rm C^{18}O \,and\, C^{17}O$ inside the disk appears different from that of CO (panels (c) and (d), Fig. \ref{abund}), while the distribution of $\ce{^{13}_{}CO}$ is similar to that of CO (panel (b), Fig. \ref{abund}). For the two less abundant isotopologues the warm molecular layer is indeed much thinner and separated by a warm molecular finger at the midplane at small radii, $R$<200 AU. This difference in the molecular distribution is the effect of the isotope selective photodissociation and it is highlighted in Fig. \ref{abu_ratio} through the abundance ratios: 
\begin{equation}
Rat^n_{xy}= \frac{n(\ce{^{x}_{}C}\ce{^{y}_{}O})[\ce{^{12}_{}C}][\ce{^{16}_{}O}]}{n(\ce{^{12}_{}CO})[\ce{^{x}_{}C}][\ce{^{y}_{}O}]}.
\end{equation}
This indicates that the conventional way of deriving the CO isotopologue abundances by simply dividing the $\ce{^{12}_{}CO}$ abundance by the isotope ratios is not accurate. 

In Figure \ref{abu_ratio} the ratios of the abundances found using these two methods are presented, i.e., the differences in the predictions whether or not the isotope-selective processes are taken into account. In the right-hand panels, the regions where $\rm C^{18}O \,and\, C^{17}O$ are not yet self-shielded, while $\rm ^{12}CO$ is, blow-up clearly. In those regions the isotopologue-ratios are lower up to a factor of 40 compared with just rescaling the isotope ratios.These isotope-selective effects for the $^{18}$O and $^{17}$O
    species are also clearly seen when the values of $Rat^n_{xy}$
    for the total number of CO molecules summed over the whole disk are
    compared in Table \ref{table:rat}.
\begin{table}[!]
\caption{Ratios of the total number of molecules $Rat^n_{xy}$ in the gas summed over the entire disk obtained using the ISO and the NOISO networks for every model.}
\label{table:rat}
\centering
\begin{tabular}{lccccc}
\toprule
& $M_{\rm d} \, [M_{\odot}]$&$f_{\rm large}$&$Rat^n_{13}$&$Rat^n_{18}$&$Rat^n_{17}$\\
\cmidrule(lr){2-3}
\cmidrule(lr){4-6}
T Tau&10$^{-4}$&10$^{-2}$& 0.95& 0.20&0.22\\
&&0.99&0.94&0.09&0.09\\
&10$^{-3}$&10$^{-2}$&0.93&0.55&0.58\\
&&0.99&0.98&0.42&0.43\\
&10$^{-2}$&10$^{-2}$&0.95&0.75&0.79\\
&&0.99&0.98&0.71&0.73\\
\cmidrule(lr){2-3}
\cmidrule(lr){4-6}
Herbig&10$^{-4}$&10$^{-2}$&0.95&0.23&0.24\\
&&0.99&0.96&0.07&0.08\\
&10$^{-3}$&10$^{-2}$&0.94&0.45&0.49\\
&&0.99&0.96&0.12&0.13\\
&10$^{-2}$&10$^{-2}$&0.94&0.77&0.84\\
&&0.99&0.98&0.64&0.66\\
\bottomrule
\end{tabular}
\end{table}

Another way to present the isotopologue fractionation is through $\mathcal{R}$, the cumulative column density ratios normalized to $\rm ^{12}CO$ and the isotopic ratios , as done e.g. by \cite{Visser09}:
\begin{equation}
\mathcal{R}(z)=\frac{N_z(\ce{^{x}_{}C}\ce{^{y}_{}O})[\ce{^{12}_{}C}][\ce{^{16}_{}O}]}{N_z(\ce{^{12}_{}CO})[\ce{^{x}_{}C}][\ce{^{y}_{}O}]},
\label{cum_dens}
\end{equation}
where [X] is the elemental abundance of isotope X and
\begin{equation}
N_z(\ce{^{x}_{}C}\ce{^{y}_{}O})(z)=\int\limits_{z}^{z_{\rm surf}} n(\ce{^{x}_{}C}\ce{^{y}_{}O}) \, \text{d}z' \, ,
\label{ccd}
\end{equation}
is the column density, integrated from the surface of the disk ($z_{\rm surf}$) down to the height $z$. 

\begin{figure}
  \resizebox{\hsize}{!}
             {\includegraphics[width=0.7\textwidth]{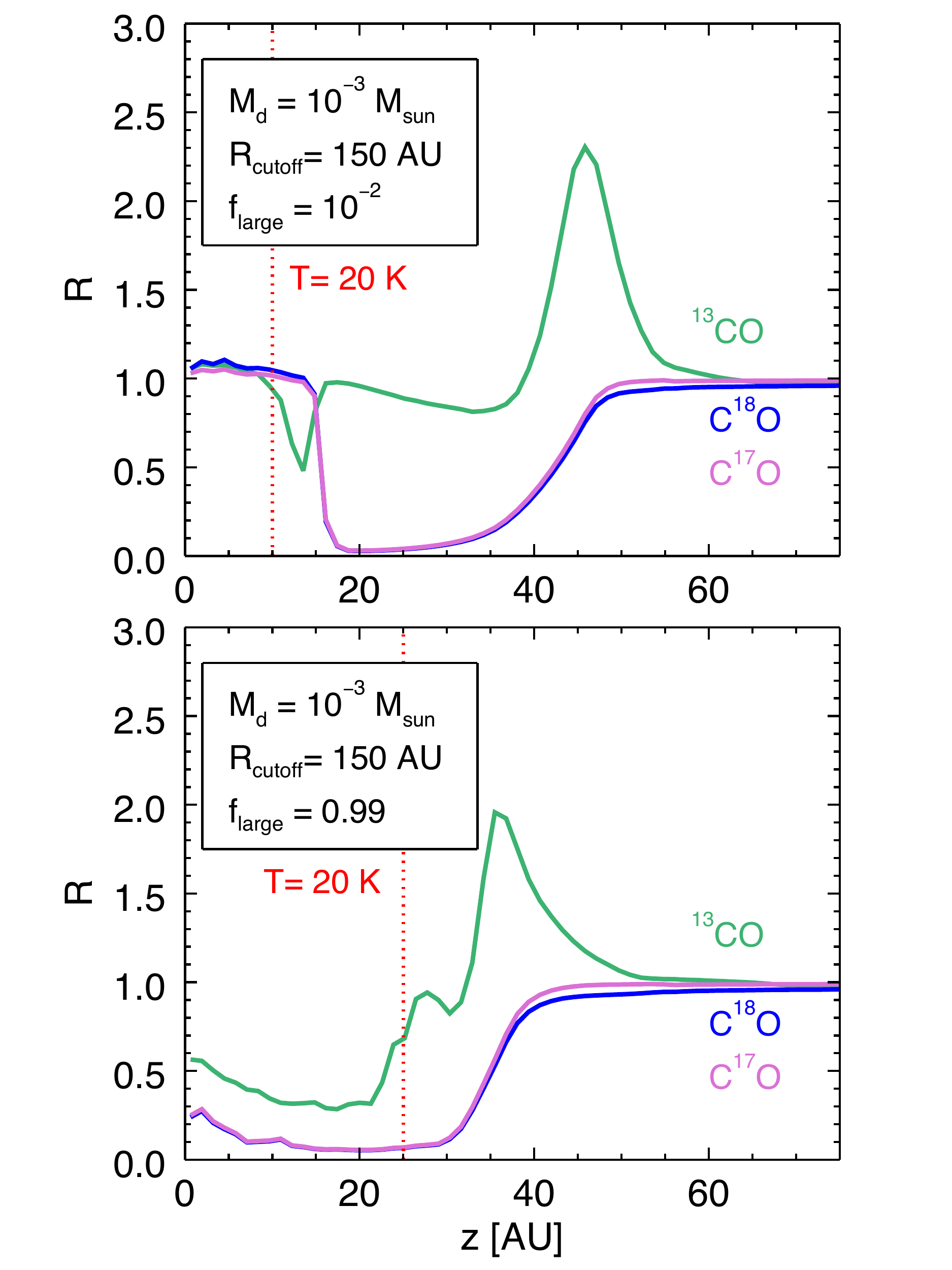}}
      \caption{Cumulative column density ratios normalized to $\rm ^{12}CO$ and the isotopic ratios (see eq. \ref{cum_dens}) shown as a function of height for a vertical cut through the disk at a radius of 150 AU. In the top panel the case of a $10^{-3} M_{\odot}$ disk around a T Tauri star with $f=10^{-2}$ (only small grains) is shown, while in the bottom panel the settling parameter is $f_{\rm large}=0.99$. The dotted red lines mark where the temperature reaches 20 K, below which CO starts to freeze out.}
       \label{cutoff}
\end{figure}

Figure \ref{cutoff} shows $\mathcal{R}$ as function of disk height through a vertical cut at a radius of 150 AU for the three isotopologues in two representative models. Consistent with \cite{Visser09} for $\rm C^{18}O \,and\, C^{17}O$ $\mathcal{R}$ is found to be constant around unity, until it dips strongly at intermediate heights. There $\mathcal{R}$ perceptibly drops because $\rm ^{12}CO$ is already self-shielded and survives the photodissociation, while UV photons still dissociate $\rm C^{18}O \,and\, C^{17}O$.  This is the region where isotope-selective effects are most detectable. If only small grains are present in the disk (upper panel of Fig. \ref{cutoff}) the temperature at which CO can freeze out ($T_{\rm dust} \lesssim 20 $ K) is reached below $z=$10 AU, where the cumulative ratios are back close to the elemental isotope ratios ($\mathcal{R}\sim 1$). On the other hand if large grains are considered (lower panel) this threshold shifts to 25 AU, just in the region where isotope-selective dissociation is most efficient. 
For heights smaller than 25 AU the tiny amount of $\rm C^{18}O \,and\, C^{17}O$ remaining in the gas phase does not add to the column density, so $\mathcal{R}$ is effectively frozen at around 0.2 for both isotopologues.
$\rm ^{13}CO$, on the other hand, has less fractionation in both models. $\mathcal{R}$ however increases by a factor of three at intermediate heights (around $z$=40 AU) for this isotopologue due to gas-phase reactions.
For each isotopologue, isotope-selective effects are maximized if mm-sized grains are present in the disk, as further discussed in section \ref{line_fluxes}. 

A negligible fraction of CO is in solid CO in our models, in
particular for the warm Herbig disks. Only the more massive cold disks
around T Tauri stars have a solid CO fraction comparable to that of
gaseous CO. Overall, a large fraction of oxygen is locked up in water
ice in the models. In particular, the excess $^{18}$O and $^{17}$O
produced by the isotope-selective photodissociation is turned into
water ice, which has a much larger binding energy and thus stays on
the grains even in warm disks. If this water ice subsequently comes
into contact with solids which drift inwards, this could be an
explanation for the anomalous $^{18}$O and $^{17}$O isotope ratios found in
meteorites \citep{Lyons05,Visser09}.

\subsection{Line fluxes}
\label{line_fluxes}
\begin{figure*}
   \resizebox{\hsize}{!}
             {\includegraphics[width=1.\textwidth]{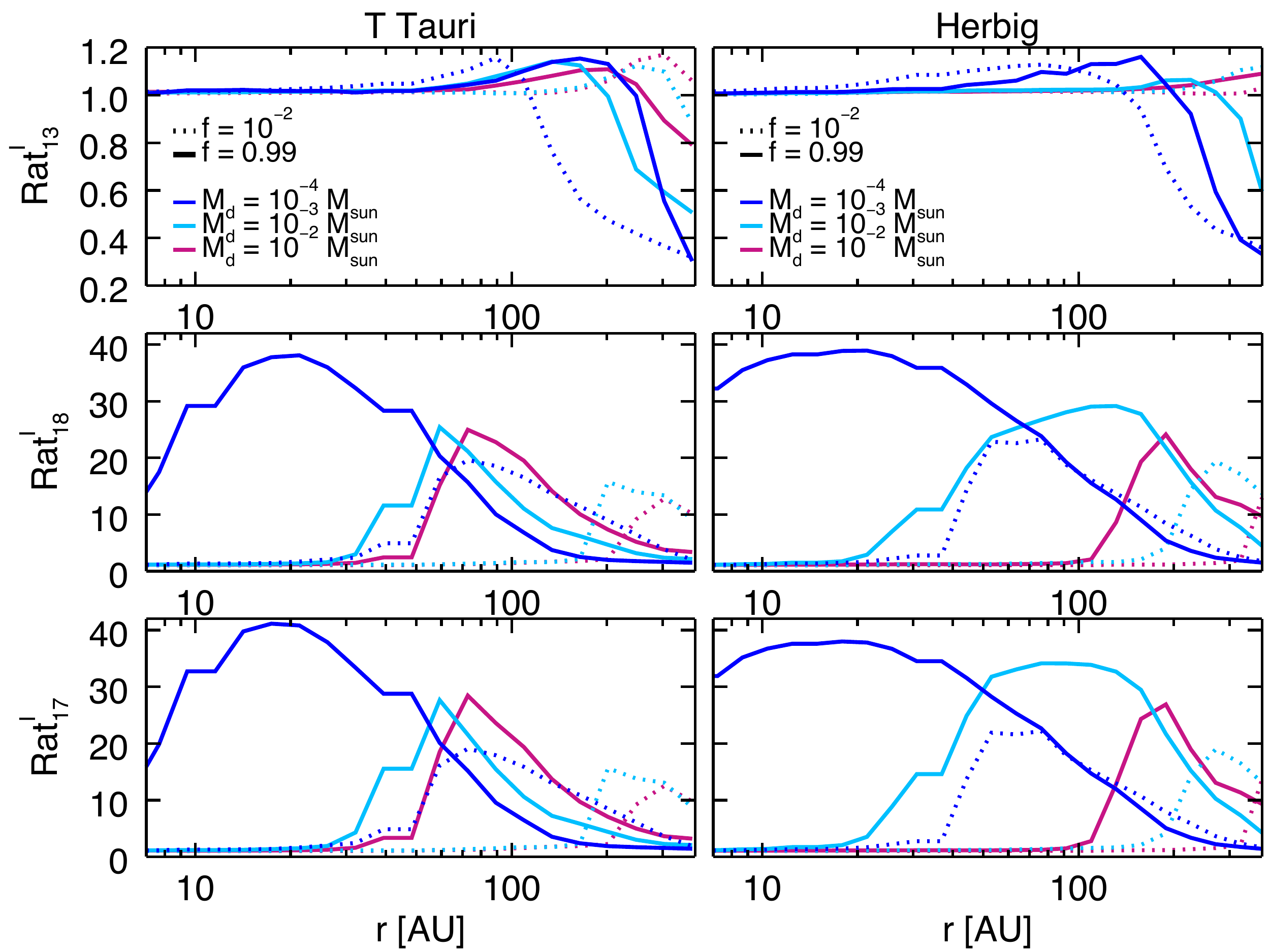}}
      \caption{Ratios of CO isotopologues line intensity ($J$=3-2) obtained with the NOISO and ISO networks as a function of the disk radius. Rat$^{\rm I}_{13}=I[\rm ^{13}CO]_{\rm NOISO}/$$I[\rm^{13}CO]_{\rm ISO}$, Rat$^{\rm I}_{18}=I[\rm C^{18}O]_{\rm NOISO}/$$I[\rm C^{18}O]_{\rm ISO}$ and Rat$^{\rm I}_{17}=I[\rm C^{17}O]_{\rm NOISO}/$$I[\rm C^{17}O]_{\rm ISO}$ were defined. Dotted lines refer to models where $f=10^{-2}$, solid lines to $f_{\rm large}=0.99$. Different colors indicate models with different disk masses: $M_{\rm d}=10^{-4} M_{\rm \odot}$ in blue, $M_{\rm d}=10^{-3} M_{\rm \odot}$ in light blue and $M_{\rm d}=10^{-2} M_{\rm \odot}$ in purple. Models with T Tauri stars are shown in the left panels, those with Herbig stars in the right panels. }
       \label{line_int}
\end{figure*}

While the 2D representations of molecular abundances shown in Figure \ref{abund} and \ref{abu_ratio} are useful to understand the chemical and physical structures of the disk, line fluxes are a better proxy for quantifying the observable effects given by the isotope-selective processes. In Fig. \ref{line_int} we present the radial intensity ratios of the $J$=3-2 line for $\rm ^{13}CO, C^{18}O \,and\, C^{17}O$ obtained with the NOISO and ISO networks. 
Spectral image cubes are obtained with DALI from the solution of the radiative transfer equation, assuming the disk is at a distance of 100 pc. The derived line intensities are not convolved with any observational beam. For $\rm ^{13}CO$ the line intensities obtained with the two networks are very similar, leading to line ratios within a factor of two. On the other hand for $\rm C^{18}O \,and\, C^{17}O$  the line intensity ratios obtained with the two networks differ by up to a factor of  40. This means that  an analysis not considering isotope-selective effects, would lead to an over prediction of  the two less abundant isotopologues line intensity at radii around 100 AU. Using the NOISO network would overpredict the $\rm C^{18}O \,and\, C^{17}O$ lines intensities, up to a factor of 40. 
\begin{figure*}
\centering
    {\includegraphics[width=0.8\textwidth]{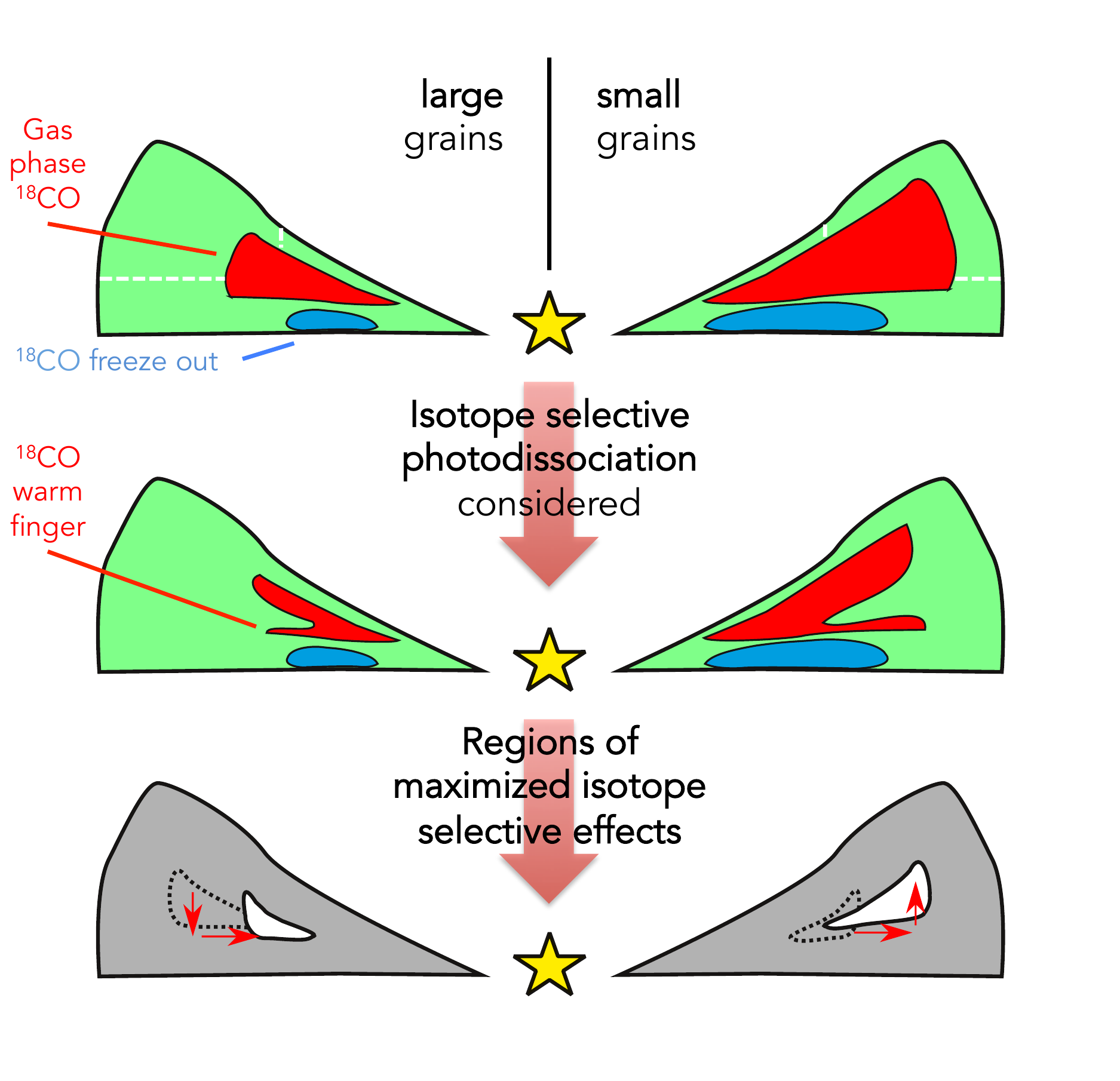}}
      \caption{Sketch of the 2D abundances of $\rm C^{18}O$ in one quadrant of the disk. On the left hand side mm-sized grains are present in the disk ($f_{\rm large}$=0.99), while on the right just (sub) $\mu$m-sized grains are considered ($f_{\rm large}$=10$^{-2}$).  In the upper panel the 2D abundance predicted by the NOISO network (i.e., not considering isotope selective photodissociation) is plotted, in the middle panel that inferred by the ISO network, i.e., with an isotopologue selective treatment of isotopologues. The red regions show where CO isotopologues are in the molecular phase, while the blue ones where they are frozen onto grains.
The regions where the two predictions differ more are those in which isotope-selective processes are maximized and are highlighted in the lower panel by the white shapes. The black dotted contours and the red arrows show the difference between ISO and NOISO. Since small grains are more efficient in shielding from photodissociating photons, $\rm C^{18}O$ can survive in the gas phase at larger heights and further out (white dotted lines in the upper panel). Accordingly, also the white regions, that show the difference between ISO and NOISO, in the lower panel shifts upward and outward in the case of small grains.}
       \label{cartoon}
\end{figure*}

\subsubsection{Dependence on the parameters}
It is interesting to check how the observables depend on the variation of the parameters listed in Section \ref{grid}. The focus is in particular to the line intensity ratios found for $\rm C^{18}O \,and\, C^{17}O$, where the effects are more prominent.

Leaving unchanged the dust composition and increasing the disk mass from $M_{\rm d}=10^{-4} M_{\rm \odot}$ to $M_{\rm d}=10^{-2} M_{\rm \odot}$, the region where the isotope-selective photodissociation is most effective shifts toward the outer regions. The peak of the intensity ratios between the ISO and NOISO networks is located there (second and third panels of Fig. \ref{line_int}). They are defined as Rat$^{\rm I}_{13}=I[\rm ^{13}CO]_{\rm NOISO}/$$I[\rm^{13}CO]_{\rm ISO}$, Rat$^{\rm I}_{18}=I[\rm C^{18}O]_{\rm NOISO}/$$I[\rm C^{18}O]_{\rm ISO}$ and Rat$^{\rm I}_{17}=I[\rm C^{17}O]_{\rm NOISO}/$$I[\rm C^{17}O]_{\rm ISO}$. At higher masses the column densities at which CO UV-lines saturate are reached earlier: CO and its isotopologues can therefore survive at larger radii and heights and the observable effects of isotope-selective photodissociation come from the outer regions. Also, higher mass disks have colder deeper regions and thus more CO freeze-out onto dust grains in the midplane. 

A similar and even more substantial shift toward larger radii is seen considering two models with the same disk mass but with different dust composition. Sub-micron-sized particles are much more efficient than mm-sized grains in shielding FUV phodissociating photons, therefore the observable effects of isotope-selective photodissociations come from outer regions. This behavior is sketched in Fig. \ref{cartoon} where it is shown that the CO emitting region shifts deeper into the disk and inward with grain growth, as found previously by \cite{Jonkheid04} and \cite{Aikawa06}.

The magnitude of the isotope-selective effects is less in the case of small grains, however. The peak in line intensity ratios is a factor of 2 lower than for large grains.
The explanation is in the location of the freeze out zone. In the case of large grains the region where less abundant CO isotopologues are highly fractionated partially coincides with the freeze out zone (see Fig. \ref{cutoff}). For $T_{\rm dust}\lesssim20$ K almost all CO molecules are frozen onto grains, except for a small fraction which is photodesorbed by FUV radiation, and the tiny fraction that remains in the gas phase can not substantially enhance the line intensity.

The last parameter to consider is the spectrum of the central star. In Fig. \ref{line_int} the line intensity ratios obtained considering a T Tauri star and a Herbig star are presented. There is no substantial difference in the peak values of the ratios, but their location depends on the stellar spectrum. For low mass disks the peak for both T Tauri and Herbig disks is in the inner region, while for high mass disks the Herbig emission peaks at larger radii. For example in the case of the model with the highest mass, $M_{\rm d}=10^{-2} M_{\rm \odot}$, and with large grains, $f_{\rm large}=0.99$, the peak shifts drastically outward at $\approx$200AU for a Herbig star. This shift is caused by the 
fact that Herbig stars have more CO dissociating UV photons ($\lambda < 1100$ \AA), compared with T Tauri stars. 

\subsection{Line optical depth}

$\rm C^{18}O$ is frequently used as a mass tracer because of its low abundance. Compared with less rare isotopologues like $\rm ^{12}CO$ and $\rm ^{13}CO$, its detection allows to probe regions closer to the midplane and thus the bulk of the disk mass. $\rm C^{18}O$ is a good mass tracer as long as its lines are optically thin and this may not always be the case. Then $\rm C^{17}O$, being rarer than $\rm C^{18}O$, can be used to probe those regions where the optical depth is greater than unity.

One way to investigate line optical thickness is to compare the line intensity radial profile with the column density radial profile. In the case of optically thin emission, they should indeed follow the same trend, since column density counts the total number of molecules at a given column. If the line intensity radial profile does not follow that of the column density, this is a signature of an opacity effect.

\begin{figure*}
   \resizebox{\hsize}{!}
             {\includegraphics[width=1.\textwidth]{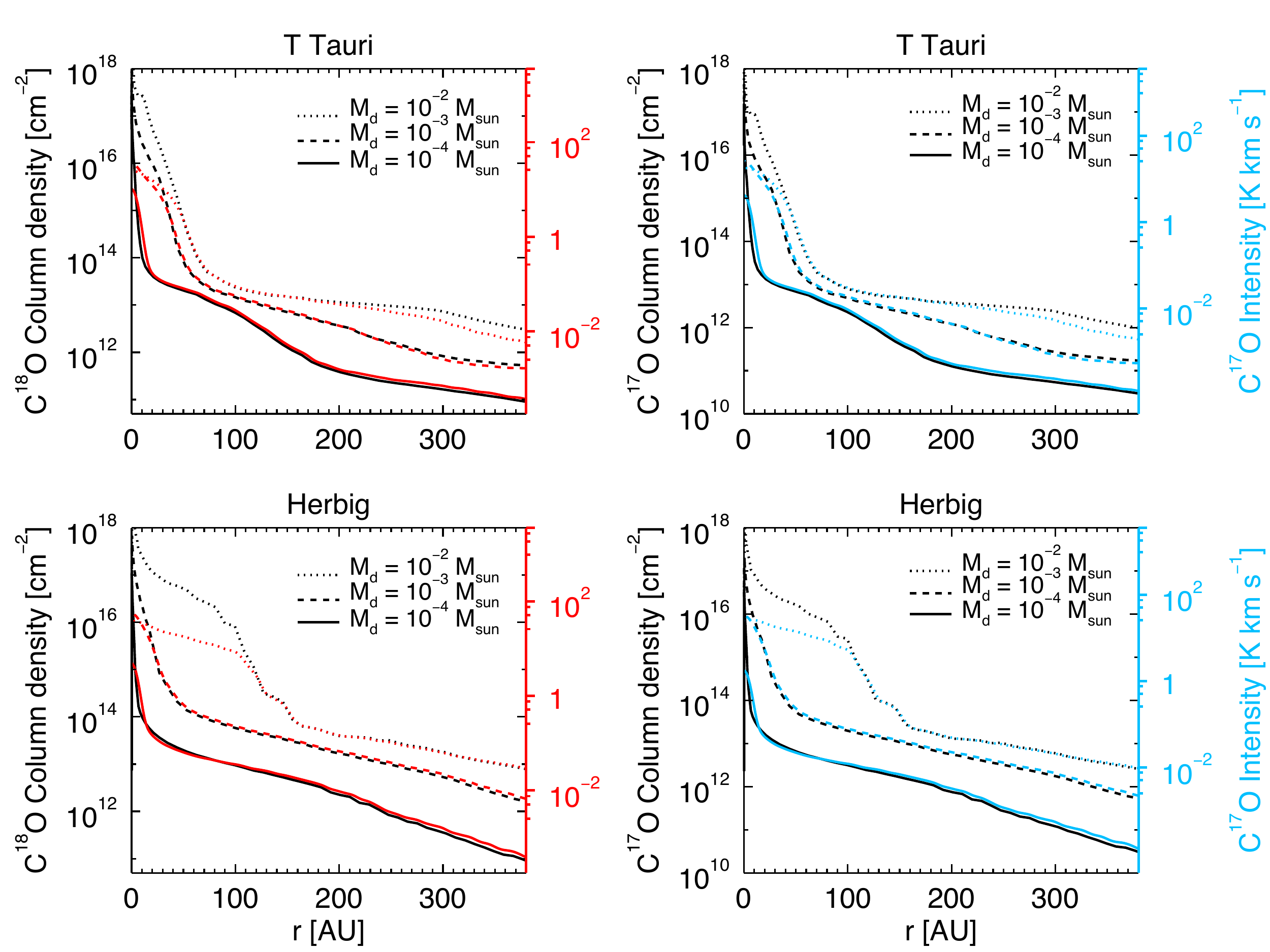}}
      \caption{Column density radial profiles of $\rm C^{18}O$ and $\rm C^{17}O$ compared with beam convolved line intensity ($J$=3-2) radial profiles. Different line styles represent models with various disk masses; column densities in black, $\rm C^{18}O$ and  $\rm C^{17}O$ line intensities in red and in blue respectively, convolved with a 0.1" beam. Models with a central T Tauri star are presented in the left panels, while those with a Herbig star in the right panels. Just models with $f_{\rm large}=0.99$ are shown.}
       \label{col_int}
\end{figure*}

In Fig. \ref{col_int} column density radial profiles are compared with line intensity radial profiles, both for models with a T Tauri star and with a Herbig star as central objects. Column densities are shown in black, while line intensities are shown in red for $\rm C^{18}O$ and in light blue for $\rm C^{17}O$. Except for the very inner and dense regions, both isotopologues are optically thin throughout the entire disk. This is true in particular for the $10 ^{-4} M_{\odot}$ disks, but not completely for the massive $10 ^{-2} M_{\odot}$ T Tauri disk. 

\begin{figure}
   \resizebox{\hsize}{!}
             {\includegraphics[width=0.7\textwidth]{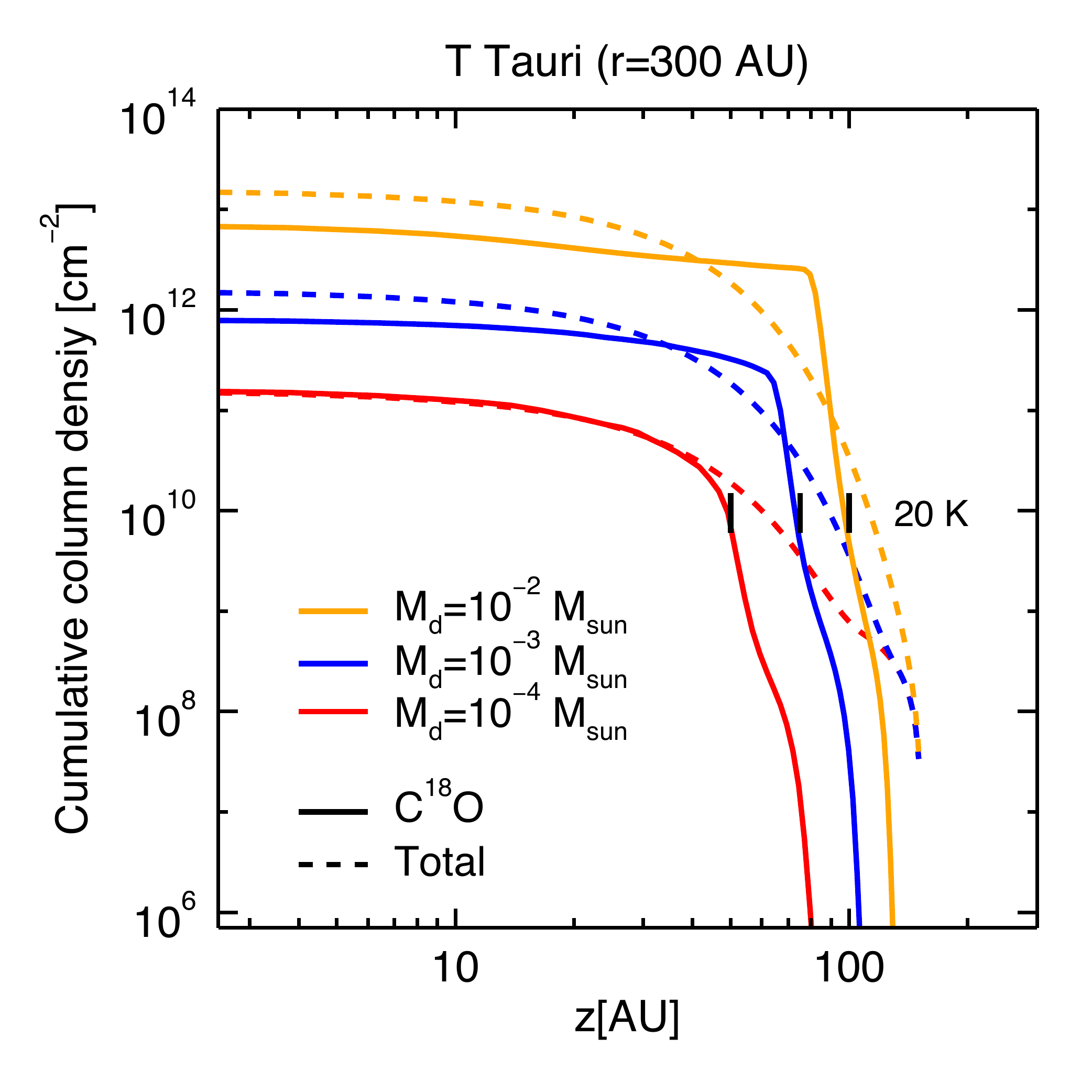}}
      \caption{Cumulative column density as a function of the disk height integrating from the surface to the midplane in a disk radial cut ($r=300$ AU). Only T Tauri models with $f_{\rm large}$=0.99 are shown. Different line colors represent models with various disk masses. Dotted lines show the scaled total gas cumulative column density (see text), while solid lines that of $\rm C^{18}O$. The black lines show where the dust temperature turns lower than 20 K for the different disk models, i.e., where CO freeze-out becomes important.}
       \label{cum_colden}
\end{figure}

It is interesting to note that in Fig. \ref{col_int} in the outer regions, i.e. for $r>200$ AU, the column densities obtained with different models do not scale linearly with mass. This depends on the amount of $\rm C^{18}O$ and $\rm C^{17}O$ frozen onto grains, which varies for the different models. In Fig. \ref{cum_colden} the $\rm C^{18}O$ and the total (H $+ 2\rm H_{2}$) cumulative column densities (eq. \ref{ccd}) are shown for a cut at $r = 300$ AU. The total cumulative column densities are divided by a factor $10^{-9}$. For $z=0$, i.e. where the column densities are integrated from the surface up to the midplane, the total cumulative column density scales linearly with mass as expected, but that of  $\rm C^{18}O$ does not, because of the different freeze-out zones.


\section{Discussion}
\label{discussion}
In order to properly compare model predictions with data, one needs to convolve output images with the same beam as used for observations. Whether or not the isotope-selective effects presented in previous sections can indeed be detected may vary with different beam sizes. Also, estimates of the total disk masses may be affected by this choice. Below we discuss this issue and provide estimates of the magnitude of this effect in determining disk masses. 

\subsection{Beam convolutions}

In order to investigate the sensitivities of mass estimates on isotope selective effects the output images obtained implementing the ISO and NOISO network have been convolved with different observational beams. Four beam sizes have been adopted for the convolution: 0.1", 0.5", 1.0", and 3.0" . The smallest are useful to simulate high resolution observations (e.g. ALMA observations), while the largest are used to compare the models with lower resolution observations (e.g SMA observations) or disk integrated (single dish) measurements.

In Fig.\ref{conv} radial distributions of $\rm C^{18}O$ line intensity obtained with different beam convolutions are presented for one of the T Tauri disk models located at 100 pc. Different colors refer to various beam sizes, while the line style shows whether isotope-selective effects are taken into account or not: solid line if they are considered (ISO network), dotted line otherwise (NOISO network). The predictions obtained with the two chemical networks always differ, but they do so in diverse ways if different beam sizes are adopted. In the extreme case of the 3" beam, the NOISO network overproduces the disk integrated $\rm C^{18}O$ line intensity by a factor of 5. On the other hand if a very small beam is used for the convolution, i.e. 0.1" beam, the line intensities are the same at very small radii for the two models; however at larger distances the NOISO network overestimates the line intensity, but not always by the same factor. Such a small beam allows the resolving of substructures that are smeared out by a larger beam. As expected, the regions of the disk where line intensity ratios of ISO and NOISO are high are the same as those for which the unconvolved line intensities obtained with the two networks mostly differ (see Fig. \ref{line_int}).

\subsection{Mass estimates}
\begin{figure*}
\centering
  \resizebox{\hsize}{!}
             {\includegraphics[width=0.9\textwidth]{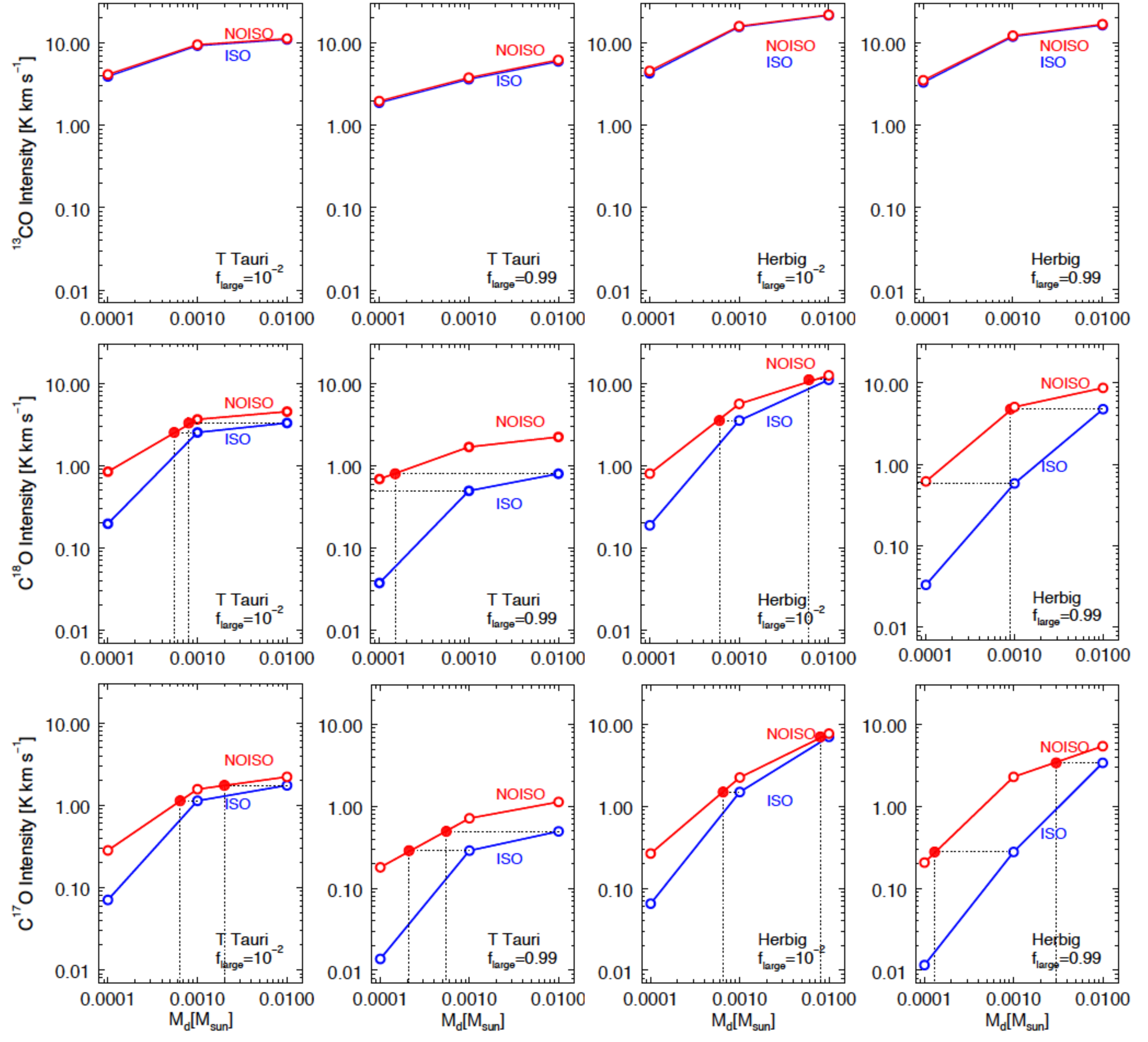}}
      \caption{Disk integrated $\rm ^{13}CO$ (top), $\rm C^{18}O$ (middle) and $\rm C^{17}O$ (bottom) line intensities ($J$=3-2) obtained with disk masses $M_{\rm d}= 10^{-4}, 10^{-3}, 10^{-2} M_{\odot}$ are shown as a function of the disk mass. The beam size is 3". The red line relates the intensity values obtained for the three disk masses with the NOISO network, while the blue line shows those with the ISO network. Dotted lines indicate by which factor the total disk mass is underestimated if no isotope-selective processes are considered. Results are presented for $f_{\rm large}=10^{-2}$, 0.99 and for T Tauri and Herbig disks.}
       \label{3arcsec}
\end{figure*}

The question to answer now is how much disk mass estimates may vary if a proper treatment of CO isotopologues is adopted. Unfortunately $\rm ^{13}CO$, the isotopologue less affected by isotopologue selective processes, becomes optically thick at substantial heights from the midplane, and is thus not a good mass tracer (Fig. \ref{3arcsec}, top). Therefore just  $\rm C^{18}O$  and  $\rm C^{17}O$ can be considered for mass estimates.  

\begin{figure}[]
 \resizebox{\hsize}{!}
             {\includegraphics[width=0.7\textwidth]{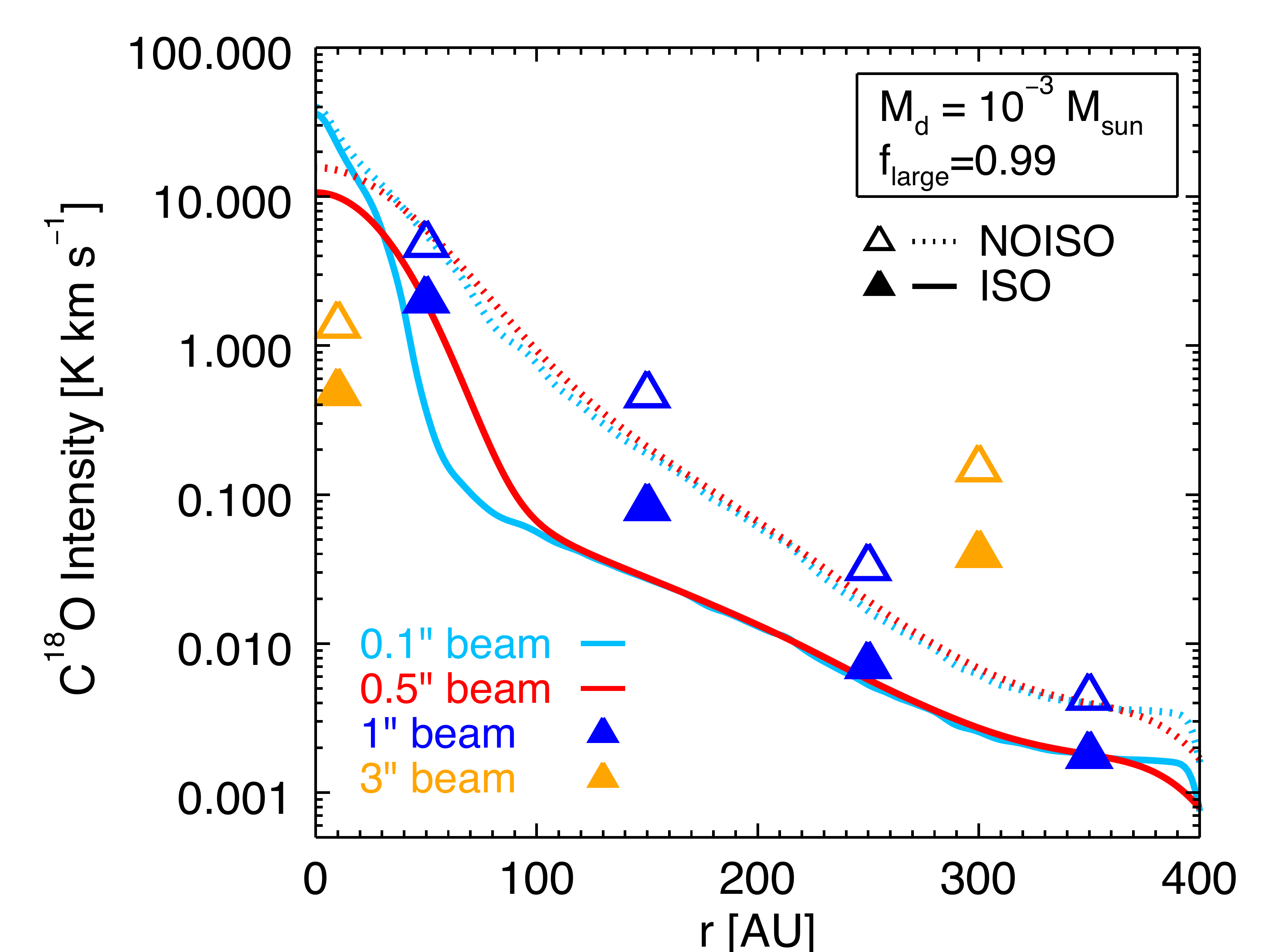}}
      \caption{Radial profiles of the $\rm C^{18}O$ line intensity ($J$=3-2) considering different observational beams for a particular disk model (T Tauri star, $M_{\rm disk}=10^{-3} M_{\odot}, f_{\rm small}=0.99$). Different colors show different choices of the observational beam size. Solid lines represent the intensities obtained considering the isotope-selective processes (ISO network), while dotted lines if an approximative analysis is carried out (NOISO network). For the 0.1" and 0.5" beam, full lines are presented, whereas for the 1" and 3" beam they are binned at the spatial resolution. For the 3" case, only the disk integrated value (or the value that just resolves the outer disk) is shown.}
       \label{conv}
\end{figure}

In the case of spatially resolved observations (0.1"-1"), the difference between the line intensities obtained with the NOISO and ISO networks varies through the disk (see Fig. \ref{conv}). This is because isotope selective processes have different influence in different regions of the disk. Therefore, as our analysis concluded, no simple scaling relation can be given, but a full thermochemical model should be run to estimate the correct disk mass.

On the other hand, in the extreme case of a 3" observational beam, where the disk substructures are not resolved, the bulk of the disk integrated line intensity is obtained. In Fig. \ref{3arcsec} the integrated line intensities obtained with disk masses $M_{\rm d}= 10^{-4}, 10^{-3}, 10^{-2} M_{\odot}$ are shown, both implementing the NOISO and the ISO network. The red line relates the intensity values obtained for the three disk masses with the NOISO network, while the blue line shows those with the ISO network. The dotted lines indicate the different masses inferred by the two networks, and thus the factor by which the total disk mass is underestimated if isotope-selective processes are neglected. Both grain growth level and variation in the stellar spectrum are investigated. The inferred masses, both considering isotope selective processes and not considering them, are presented in Table \ref{tab:mass}, together with their ratios. 
In practice, a given $\rm C^{18}O$ and $\rm C^{17}O$ line intensity is observed and plots such as those in Fig. \ref{3arcsec} can then be used to draw a horizontal line and read off the disk masses.

The grid of models presented in this paper explores only a few parameters and it is not possible to draw a general correction rule; we stress that the values in Table \ref{tab:mass} are only indicative and may well reproduce extreme cases. On the other hand it is possible to find trends in the results. 
For a given $\rm C^{18}O$ or $\rm C^{17}O$ intensity , the total disk mass is always underestimated if the NOISO network is used, i.e. if isotope-selective effects are not properly considered. The underestimates are larger if mm-sized grains are present in the disk (i.e., where photodissociation is most efficient) and for a T Tauri star as central object, for which the disk is cold. Moreover, the results differ by a larger factor if more massive disks are considered, i.e. disks with masses that range from  $10^{-3} M_{\odot}$ to $10^{-2} M_{\odot}$. 

\cite{Williams14} used a simple parametric
approach to infer disk masses from their $^{13}$CO and C$^{18}$O
data assuming constant isotopologue abundance ratios in those
regions of the disk where CO is not photodissociated or frozen out.
Effects of isotope selective photodissociation are treated by
decreasing the C$^{18}$O/CO abundance ratio by an additional constant factor
of 3 throughout the disk. Although our mini-grid is much smaller
than that of \cite{Williams14}, we tend to find higher $^{13}$CO and
lower C$^{18}$O intensities by up to a factor of a few for the same
disk mass, especially for the models with large grains at the low
mass end. A more complete parameter study of our full chemical
models is needed for proper comparison.

\begin{table*}[!]
\caption{Disk mass estimates obtained from a given observation of $\rm C^{18}O$ or $\rm C^{17}O$ neglecting isotope selective photodissociation, then considering it. The inferred disk masses are obtained with the disk integrated line intensities given as outputs with the NOISO and ISO networks (see Fig. \ref{3arcsec}). The ratios between these two estimates are also reported for all the disk models. Those are the correction factors that should be applied to mass estimates, obtained ignoring isotope-selective photodissociation. }
\label{tab:mass}
\centering
\begin{tabular}{ccccccc}
\toprule
& \multicolumn{2}{c}{True $M_{\rm d} \, [M_{\odot}]$}&\multicolumn{2}{c}{Inferred NOISO $M_{\rm d} \, [M_{\odot}]$} & \multicolumn{2}{c}{Ratio$ (M_{\rm True}/M_{\rm NOISO})$}  \\
\cmidrule(lr){2-3}
\cmidrule(lr){4-5}
\cmidrule(lr){6-7}
& small & large & small & large & small & large \\
\midrule
&\multicolumn{2}{c}{T Tauri}& \multicolumn{2}{c}{T Tauri}&\multicolumn{2}{c}{T Tauri}\\
\cmidrule(lr){2-3}
\cmidrule(lr){4-5}
\cmidrule(lr){6-7}
$\ce{C^{18}O}$&$10^{-3}$&$10^{-3}$&$5.5\cdot10^{-4}$&$5\cdot10^{-5}$&1.8&20\\
&$10^{-2}$&$10^{-2}$&$8\cdot10^{-4}$&$1.5\cdot10^{-4}$&$12.5$&67\\
\cmidrule(lr){2-3}
\cmidrule(lr){4-5}
\cmidrule(lr){6-7}
$\ce{C^{17}O}$&$10^{-3}$&$10^{-3}$&$6.4\cdot10^{-4}$&$2.1\cdot10^{-4}$&1.6&4.8\\
&$10^{-2}$&$10^{-2}$&$2\cdot10^{-3}$&$5.5\cdot10^{-4}$&5&20\\
\midrule
&\multicolumn{2}{c}{Herbig}& \multicolumn{2}{c}{Herbig}&\multicolumn{2}{c}{Herbig}\\
\cmidrule(lr){2-3}
\cmidrule(lr){4-5}
\cmidrule(lr){6-7}
$\ce{C^{18}O}$&$10^{-3}$&$10^{-3}$&$6\cdot10^{-4}$&$10^{-4}$&1.7&10\\
&$10^{-2}$&$10^{-2}$&$6\cdot10^{-3}$&$9\cdot10^{-4}$&1.7&$11$\\
\cmidrule(lr){2-3}
\cmidrule(lr){4-5}
\cmidrule(lr){6-7}
$\ce{C^{17}O}$&$10^{-3}$&$10^{-3}$&$6.5\cdot10^{-4}$&$1.3\cdot10^{-4}$&1.5&7.8\\
&$10^{-2}$&$10^{-2}$&$8\cdot10^{-3}$&3$\cdot10^{-3}$&1.3&3.3\\
\bottomrule
\end{tabular}
\end{table*}

\subsection{The case of TW Hya}
TW Hya is one of the best studied protoplanetary disks, being the closest T Tauri system, at a distance of 54$\pm$6 pc from the Earth \citep{vl07}. Also, it is the only case in which the fundamental rotational transition of HD has been detected with the \emph{Herschel Space Observatory} \citep{Bergin13}. Observations of HD emission lines provide an independent disk mass determination, assuming that the HD distribution follows that of $\rm H_2$. From these observations \cite{Bergin13} determine a total disk mass larger than $5\cdot10^{-2} M_{\odot}$. \cite{Favre13} report also spatial integrated $\rm C^{18}O$ ($2 -1$) observations of TW Hya with SMA. From their analysis an underestimate of the disk mass by a factor between 3 and $100$ is found. In their modeling they do not treat isotope selective photodissociation in a self-consistent way and conclude instead that the carbon abundance is low.

For the model in our grid that best represents the
TW Hya characteristics (T Tauri star, large grains, $ M_{\rm d} = 10^{-2} M_{\odot}$),
a mass correction factor of at least 20 is found (see
Table \ref{tab:mass}). This demonstrates that accounting for isotope selective
processes could mitigate the disagreement in mass determinations. The
spatial resolution of the SMA data of 100--150 AU is just in the
regime where the effects are maximal. More accurate modeling of TW
Hya implementing the full isotope selective photodissociation using
the same disk model as in \cite{Favre13}, \cite{Bergin13} and \cite{Cleeves14} with the observed stellar UV spectrum is needed to support this claim.

%__________________________________________________________________

\section{Summary and Conclusions}
\label{Summary}
This paper presents a detailed treatment of CO isotope selective photodissociation in a complete disk model for the first time. A full thermo-chemical model is used, in which less abundant CO isotopologues have been added into the chemical network as independent species and the corresponding self-shielding factors are implemented. Abundances and line intensities are obtained as outputs for a grid of disk models, with and without isotope-selective photodissociation. The main conclusions are listed below.
\begin{itemize}
\item If CO isotope selective photodissociation is considered, the abundances of CO isotopologues are affected. In particular there are regions in the disk where $\rm C^{18}O$ and  $\rm C^{17}O$ show an underabundance with the respect to  $\rm ^{12}CO$, when compared with the overall elemental abundance ratios.
\item The CO isotopologues line intensity ratios vary if isotope selective photodissociation is properly considered. The line intensity ratios are overestimated up to a factor of 40 at certain disk radii, if isotope selective processes are not included into the modeling. 
\item A consequence of these results is that the disk mass can be
underestimated by up to almost two orders of magnitude if a single line is
observed and isotope selective effects are not properly taken into
account.  The effects are largest for cold disks with large grains
where the isotope selective effects are maximal close to the
freeze-out zone, i.e., the same region of the disk from which most of
the $\rm C^{18}O$ and $ \rm C^{17}O$ emission arises. Also, the situation is worse for
single-dish data or for low spatial resolution interferometry which
just resolves the outer disk.
\item A preliminary comparison has been done between the results presented here and the case of TW Hya. The discrepancy in mass determination observed for this object may be explained by implementing isotope selective photodissociation in a self-consistent manner. More detailed modeling of the object is however still needed.
\end{itemize}

In the future a more detailed modeling of TW Hya and other disks will be carried out. Also, a wider set of parameters will be explored in a larger grid of models, in order to provide mass estimates corrections for all kinds of disks when multiple lines are observed.

\section*{Acknowledgements}

The authors are grateful to Ruud Visser, Ilsedore Cleeves, Catherine Walsh, Mihel Kama
and Leonardo Testi for useful discussions and comments.
Astrochemistry in Leiden is supported by the Netherlands Research
School for Astronomy (NOVA), by a Royal Netherlands Academy of Arts
and Sciences (KNAW) professor prize, and by the European Union A-ERC
grant 291141 CHEMPLAN.

%___________________________________________________________________________

%__________________________________________________________________

\clearpage
\begin{appendix}

%%_ _ _ _ _ _ _ _ _ _ _ _ _ _ _ _ _ _ _ _ _ _ _ _ _ _ _ _ _
\section{Appendix A}
\label{AppendixA}
\begin{table*}[]
\caption{Species contained in the ISO chemical network (see Sect. \ref{chem_net}). \emph{Notation}: H$_2^*$ refers to vibrationally excited H$_2$;  PAH$^0$, PAH$^+$ and PAH$^-$ are neutral, positively and negatively charged PAHs, while PAH:H denotes hydrogenate PAHs; JX refers to species frozen-out onto dust grainis.}
\label{tab:chemspec}
\centering
{\footnotesize
\begin{tabular}{cccccccccc}
\hline\hline
H                               & He                              & C                               & $^{13}$C                        & N                               & O                               
&$^{17}$O                        & $^{18}$O                        & Mg                              & Si                              \\ 
S                               & Fe                              &H$_2$                           & H$_2^*$                         & CH                              & $^{13}$CH                       & CH$_2$                         
 & $^{13}$CH$_2$                   &NH                              & CH$_3$                          \\
$^{13}$CH$_3$                   & NH$_2$                          & CH$_4$                          & $^{13}$CH$_4$                   &OH                              & $^{17}$OH                       & $^{18}$OH                       & NH$_3$                          & H$_2$O                          & H$_2^{17}$O                     \\
H$_2^{18}$O                     & CO                              & C$^{17}$O                       & C$^{18}$O                       & $^{13}$CO                       & $^{13}$C$^{17}$O                
&$^{13}$C$^{18}$O                & HCN                             & H$^{13}$CN                      & HCO                             \\
HC$^{17}$O                      & HC$^{18}$O                      &H$^{13}$CO                      & H$^{13}$C$^{17}$O               & H$^{13}$C$^{18}$O               & NO                              & N$^{17}$O                       & N$^{18}$O                       &H$_2$CO                         & H$_2$C$^{17}$O                  \\
H$_2$C$^{18}$O                  & H$_2^{13}$CO                    & H$_2^{13}$C$^{17}$O             & H$_2^{13}$C$^{18}$O             &O$_2$                           & O$^{17}$O                       
& $^{17}$O$_2$                    & O$^{18}$O                       & $^{17}$O$^{18}$O                & $^{18}$O$_2$                    \\
HS                              & H$_2$S                          & CO$_2$                          & CO$^{17}$O                      & C$^{17}$O$_2$                   & CO$^{18}$O                      
&C$^{17}$O$^{18}$O               & C$^{18}$O$_2$                   & $^{13}$CO$_2$                   & $^{13}$CO$^{17}$O               \\
$^{13}$C$^{17}$O$_2$            & $^{13}$CO$^{18}$O               &$^{13}$C$^{17}$O$^{18}$O        & $^{13}$C$^{18}$O$_2$            & SO                              & S$^{17}$O                       
& S$^{18}$O                       & OCS                             &$^{17}$OCS                      & $^{18}$OCS                      \\ 
O$^{13}$CS                      & $^{17}$O$^{13}$CS               & $^{18}$O$^{13}$CS               & CN                              &$^{13}$CN                       & N$_2$                           & SiH                             & CS                              & $^{13}$CS                       & HCS                             \\
H$^{13}$CS                      & SO$_2$                          & SO$^{17}$O                      & S$^{17}$O$_2$                   & SO$^{18}$O                      & S$^{17}$O$^{18}$O               
&S$^{18}$O$_2$                   & SiO                             & Si$^{17}$O                      & Si$^{18}$O                      \\
H$_2$CS                         & H$_2^{13}$CS                    &H$^+$                           & H$^-$                           & H$_2^+$                         & H$_3^+$                         & He$^+$                          
& HCO$^+$                         &HC$^{17}$O$^+$                  & HC$^{18}$O$^+$                  \\ 
H$^{13}$CO$^+$                  & H$^{13}$C$^{17}$O$^+$           & H$^{13}$C$^{18}$O$^+$           & C$^+$                           &$^{13}$C$^+$                    & CH$^+$                         
&$^{13}$CH$^+$                   & N$^+$                           & CH$_2^+$                        & $^{13}$CH$_2^+$                 \\
NH$^+$                          & CH$_3^+$                        & $^{13}$CH$_3^+$                 & O$^+$                           & $^{17}$O$^+$                    & $^{18}$O$^+$                    
&NH$_2^+$                        & CH$_4^+$                        & $^{13}$CH$_4^+$                 & OH$^+$                          \\ 
$^{17}$OH$^+$                   & $^{18}$OH$^+$                   &NH$_3^+$                        & CH$_5^+$                        & $^{13}$CH$_5^+$                 & H$_2$O$^+$                      
&H$_2^{17}$O$^+$                 & H$_2^{18}$O$^+$                 &H$_3$O$^+$                      & H$_3^{17}$O$^+$                 \\ 
H$_3^{18}$O$^+$                 & Mg$^+$                          & CN$^+$                          & $^{13}$CN$^+$                   &HCN$^+$                         & H$^{13}$CN$^+$                  & Si$^+$                          & CO$^+$                          & C$^{17}$O$^+$                   & C$^{18}$O$^+$                   \\
$^{13}$CO$^+$                   & $^{13}$C$^{17}$O$^+$            & $^{13}$C$^{18}$O$^+$            & HCNH$^+$                        & H$^{13}$CNH$^+$                 & SiH$^+$                         
&NO$^+$                          & N$^{17}$O$^+$                   & N$^{18}$O$^+$                   & SiH$_2^+$                       \\
S$^+$                           & O$_2^+$                         &O$^{17}$O$^+$                   & $^{17}$O$_2^+$                  & O$^{18}$O$^+$                   & $^{17}$O$^{18}$O$^+$            & $^{18}$O$_2^+$                  & HS$^+$                          &H$_2$S$^+$                      & H$_3$S$^+$                      \\ 
SiO$^+$                         & Si$^{17}$O$^+$                  & Si$^{18}$O$^+$                  & CS$^+$                          &$^{13}$CS$^+$                   & CO$_2^+$                        & CO$^{17}$O$^+$                  & C$^{17}$O$_2^+$                 & CO$^{18}$O$^+$                  & C$^{17}$O$^{18}$O$^+$           \\
C$^{18}$O$_2^+$                 & $^{13}$CO$_2^+$                 & $^{13}$CO$^{17}$O$^+$           & $^{13}$C$^{17}$O$_2^+$          & $^{13}$CO$^{18}$O$^+$           
& $^{13}$C$^{17}$O$^{18}$O$^+$      &$^{13}$C$^{18}$O$_2^+$          & HCS$^+$                         & H$^{13}$CS$^+$                  & SO$^+$                         \\ 
S$^{17}$O$^+$                   & S$^{18}$O$^+$                   &Fe$^+$                          & SO$_2^+$                        & SO$^{17}$O$^+$                  & S$^{17}$O$_2^+$                 
& SO$^{18}$O$^+$                  & S$^{17}$O$^{18}$O$^+$           &S$^{18}$O$_2^+$                 & HSO$_2^+$                       \\ 
HSO$^{17}$O$^+$                 & HS$^{17}$O$_2^+$                & HSO$^{18}$O$^+$                 & HS$^{17}$O$^{18}$O$^+$          
&HS$^{18}$O$_2^+$                & SiOH$^+$                        & Si$^{17}$OH$^+$                 & Si$^{18}$OH$^+$                 & H$_2$CS$^+$                     & H$_2^{13}$CS$^+$                \\
H$_3$CS$^+$                     & H$_3^{13}$CS$^+$                & HSO$^+$                         & HS$^{17}$O$^+$                  & HS$^{18}$O$^+$                  & OCS$^+$                         
&$^{17}$OCS$^+$                  & $^{18}$OCS$^+$                  & O$^{13}$CS$^+$                  & $^{17}$O$^{13}$CS$^+$           \\
$^{18}$O$^{13}$CS$^+$           & HOCS$^+$                        &H$^{17}$OCS$^+$                 & H$^{18}$OCS$^+$                 & HO$^{13}$CS$^+$                 & H$^{17}$O$^{13}$CS$^+$          & H$^{18}$O$^{13}$CS$^+$          & S$_2^+$                         &HN$_2^+$                        & HS$_2^+$                        \\ 
e$^-$                           & PAH$^0$                         & PAH$^+$                         & PAH$^-$                         
&PAH:H                           & JC                              & J$^{13}$C                       & JN                              & JO                              & J$^{17}$O                       \\
J$^{18}$O                       & JCH                             & J$^{13}$CH                      & JCH$_2$                         & J$^{13}$CH$_2$                  & JNH                             
&JCH$_3$                         & J$^{13}$CH$_3$                  & JNH$_2$                         & JCH$_4$                         \\
J$^{13}$CH$_4$                  & JOH                             &J$^{17}$OH                      & J$^{18}$OH                      & JNH$_3$                         & JH$_2$O                         & JH$_2^{17}$O                    & JH$_2^{18}$O                    &JCO                             & JC$^{17}$O                      \\ 
JC$^{18}$O                      & J$^{13}$CO                      & J$^{13}$C$^{17}$O               & J$^{13}$C$^{18}$O               
&JCO$_2$                         & JCO$^{17}$O                     & JC$^{17}$O$_2$                  & JCO$^{18}$O                     & JC$^{17}$O$^{18}$O              & JC$^{18}$O$_2$                  \\
J$^{13}$CO$_2$                  & J$^{13}$CO$^{17}$O              & J$^{13}$C$^{17}$O$_2$           & J$^{13}$CO$^{18}$O              & J$^{13}$C$^{17}$O$^{18}$O       & J$^{13}$C$^{18}$O$_2$   &&&&        \\
\hline
\end{tabular}
}
\end{table*}
\end{appendix}

\end{document}